\PassOptionsToPackage{table,dvipsnames}{xcolor}
\PassOptionsToPackage{hyphens}{url}
\documentclass[lettersize,journal]{IEEEtran}

\usepackage{amsmath,amssymb}
\usepackage{xifthen}
\usepackage{mathtools}
\usepackage{enumerate}
\usepackage{microtype}
\usepackage{xspace}
\usepackage{bm}
\usepackage[T1]{fontenc}
\usepackage{fancyhdr}
\usepackage{lastpage}
\usepackage{bbm}

\newcommand{\mbs}[1]{\pmb{#1}}
\newcommand{\vect}[1]{{\lowercase{\mbs{#1}}}}
\newcommand{\mat}[1]{{\uppercase{\mbs{#1}}}}

\newcommand{\T}{{\scriptscriptstyle\mathsf{T}}}

\renewcommand{\Re}[1][]{\ifthenelse{\isempty{#1}}{\operatorname{Re}}{\operatorname{Re}\left(#1\right)}}
\renewcommand{\Im}[1][]{\ifthenelse{\isempty{#1}}{\operatorname{Im}}{\operatorname{Im}\left(#1\right)}}

\newcommand{\av}{\vect{a}}

\newcommand{\ev}{\vect{e}}

\newcommand{\ellv}{\vect{\ell}}

\newcommand{\pv}{\vect{p}}

\newcommand{\xv}{\vect{x}}

\newcommand{\Cr}{\mathrm{C}}

\newcommand{\Er}{\mathrm{E}}
\newcommand{\Fr}{\mathrm{F}}

\newcommand{\Nr}{\mathrm{N}}

\newcommand{\Rr}{\mathrm{R}}
\newcommand{\Sr}{\mathrm{S}}
\newcommand{\Tr}{\mathrm{T}}
\newcommand{\Ur}{\mathrm{U}}
\newcommand{\Vr}{\mathrm{V}}

\newcommand{\piv}{\vect{\pi}}
\newcommand{\tauv}{\vect{\tau}}
\newcommand{\omegav}{\vect{\omega}}

\newcommand{\Pim}{\pmb{\Pi}}

\newcommand{\Lm}{\mat{l}}

\newcommand{\Xm}{\mat{x}}
\newcommand{\Ym}{\mat{y}}
\newcommand{\Zm}{\mat{z}}

\newcommand{\Ac}{{\mathcal A}}

\newcommand{\Lc}{{\mathcal L}}

\newcommand{\Nc}{{\mathcal N}}

\newcommand{\Uc}{{\mathcal U}}

\newcommand{\EE}{\mathbb{E}}

\newcommand{\RR}{\mathbb{R}}

\newcommand{\Is}{\mat{\mathsf{I}}}
\newcommand{\Ts}{\mat{\mathsf{T}}}

\newcommand{\Ps}{\mat{\mathsf{P}}}
\newcommand{\Qs}{\mat{\mathsf{Q}}}
\newcommand{\Ms}{\mat{\mathsf{M}}}

\newcommand{\CN}[1][]{\ifthenelse{\isempty{#1}}{\mathcal{N}_{\mathbb{C}}}{\mathcal{N}_{\mathbb{C}}\left(#1\right)}}

\renewcommand{\P}[1][]{\ifthenelse{\isempty{#1}}{\mathbb{P}}{\mathbb{P}\left(#1\right)}}
\newcommand{\E}[1][]{\ifthenelse{\isempty{#1}}{\mathbb{E}}{\mathbb{E}\left[#1\right]}}
\newcommand{\I}[1][]{\ifthenelse{\isempty{#1}}{\mathbb{I}}{\mathbb{I}\left\{#1\right\}}}
\renewcommand{\det}[1][]{\ifthenelse{\isempty{#1}}{\mathrm{det}}{\mathrm{det}\left(#1\right)}}
\newcommand{\trace}[1][]{\ifthenelse{\isempty{#1}}{{\rm tr}}{\mathrm{tr}\left(#1\right)}}
\newcommand{\rank}[1][]{\ifthenelse{\isempty{#1}}{\mathrm{rank}}{\mathrm{rank}\left(#1\right)}}
\newcommand{\diag}[1][]{\ifthenelse{\isempty{#1}}{\mathrm{diag}}{\mathrm{diag}\left(#1\right)}}
\newcommand{\Cov}[1][]{\ifthenelse{\isempty{#1}}{\mathsf{Cov}}{\mathsf{Cov}\left(#1\right)}}

\newcommand{\defeq}{\triangleq}

\newtheorem{remark}{Remark}
\newtheorem{theorem}{Theorem}
\newtheorem{lemma}{Lemma}
\newtheorem{simplification}{Simplification}

\newcounter{enumi_saved}
\usepackage{answers}
\Newassociation{solution}{Solution}{solutionfile}

\AtBeginDocument{\Opensolutionfile{solutionfile}[\jobname]}
\AtEndDocument{\Closesolutionfile{solutionfile}\clearpage
}

\IfFileExists{MinionPro.sty}{
}{
}

\usepackage{subcaption}

\usepackage{graphicx,psfrag}
\usepackage{multirow}
\usepackage{array}
\usepackage{setspace}
\usepackage{bbm,mathtools}
\usepackage{amssymb}
\usepackage{color}
\usepackage{enumitem}
\usepackage{stfloats}
\usepackage[flushleft]{threeparttable}
\usepackage{pgfplots}
\usepackage{booktabs}
\usepackage{makecell}
\usepackage[hidelinks]{hyperref}
\usepackage[hyphenbreaks]{breakurl}
\usepackage{cite}
\usepackage[table,dvipsnames]{xcolor}
\usepackage{xcolor}
\pgfplotsset{minor grid style={dotted}}
\pgfplotsset{major grid style={dashed}}
\usepackage{tikz}
\usetikzlibrary{fit,positioning,arrows,shapes,shapes.multipart,calc,arrows.meta,shapes.geometric,shapes.misc}
\usetikzlibrary {decorations.fractals,spy} 
\usetikzlibrary{decorations.pathreplacing,angles,quotes,calligraphy}
\usetikzlibrary{automata,fit,positioning,arrows,shapes,shapes.multipart,calc,arrows.meta}
\usetikzlibrary{plotmarks}
\usepgfplotslibrary{patchplots}
\usepackage[ruled,vlined,linesnumbered]{algorithm2e}
\usepackage[toc,acronym]{glossaries}

\makeindex

\allowdisplaybreaks

\usepackage{grffile}
\pgfplotsset{compat=newest}
\usetikzlibrary{plotmarks}
\usetikzlibrary{arrows.meta}
\usepgfplotslibrary{patchplots}

\makeatletter
\DeclareRobustCommand{\cev}[1]{%
	{\mathpalette\do@cev{#1}}%
}
\newcommand{\do@cev}[2]{%
	\vbox{\offinterlineskip
		\sbox\z@{$\m@th#1 x$}%
		\ialign{##\cr
			\hidewidth\reflectbox{$\m@th#1\vec{}\mkern4mu$}\hidewidth\cr
			\noalign{\kern-\ht\z@}
			$\m@th#1#2$\cr
		}%
	}%
}
\makeatother

\newcommand{\of}[1]{^{(#1)}}

\newcommand{\ind}[1]{{\mathbbm{1}{\{#1\}}}}

\renewcommand{\E}[2][]{\mathbb{E}_{#1}\!\left[#2\right]}
\renewcommand{\P}[2][]{\mathbb{P}_{#1}\!\left[#2\right]}

\newcommand{\sub}[1]{_{\mathrm{#1}}}

\renewcommand{\defeq}{=}

\newcommand{\bt}{{b_{\rm t}}}

\newcommand{\refresh}{{\Sr}}
\newcommand{\fail}{{\Fr}}

\newcommand{\EHrate}{{\gamma}}

\DeclareMathOperator*{\argmax}{arg\,max}
\DeclareMathOperator*{\argmin}{arg\,min}

\newcommand{\revise}[1]{{#1}} 
\newcommand{\revisee}[1]{{#1}} 

\makeatletter
\newcommand{\removelatexerror}{\let\@latex@error\@gobble}
\makeatother

\title{Timely Status Updates in Slotted ALOHA Networks With Energy Harvesting}

\author{\IEEEauthorblockN{Khac-Hoang Ngo, \emph{Member, IEEE}, Giuseppe Durisi, \emph{Senior Member, IEEE}, Andrea Munari, \emph{Senior Member, IEEE}, Francisco L\'azaro, \emph{Senior Member, IEEE}, and Alexandre Graell i Amat, \emph{Senior Member, IEEE}} 
	\thanks{\revise{Khac-Hoang Ngo is with the Department of Electrical Engineering, Linköping University, 58183 Linköping, Sweden~(email: {\tt khac-hoang.ngo@liu.se}).} Giuseppe Durisi and Alexandre Graell i Amat are with the Department of Electrical Engineering, Chalmers University of Technology, 41296 Gothenburg, Sweden~(e-mails: {\tt \{durisi, alexandre.graell\}@chalmers.se}). Andrea Munari and Francisco L\'azaro are with the Institute for Communications and Navigation, German Aerospace Center (DLR), 82234 We{\ss}ling, Germany
		(e-mail: {\tt \{andrea.munari, francisco.lazaroblasco\}@dlr.de}).}
	\thanks{Khac-Hoang Ngo has received funding from the European Union’s Horizon 2020 research and innovation programme under the Marie Sklodowska-Curie grant agreement No 101022113, and from the Excellence Center at Linköping – Lund in Information Technology (ELLIIT). Giuseppe Durisi has received funding from the Swedish Research Council under grant 2021-04970. \revise{A. Munari acknowledges the financial support of the Federal Ministry of Education and Research of Germany in the programme of ``Souverän. Digital. Vernetzt.'' Joint project 6G-RIC, project identification number: 16KISK022.}
		Alexandre Graell i Amat has received funding from the Swedish Research Council (VR) under grant 2020-03687.}
	\thanks{This paper was presented in part at the IEEE Global Communications Conference (GLOBECOM), Kuala Lumpur, Malaysia, 2023~\cite{Hoang2023Globecom}.}
}

\newacronym{AWGN}{AWGN}{additive white Gaussian noise}
\newacronym{MAC}{MAC}{multiple access channel}
\newacronym{UMRA}{UMRA}{unsourced massive random access}
\newacronym{SIMO}{SIMO}{single-input multiple-output}
\newacronym{SISO}{SIMO}{single-input single-output}
\newacronym{iid}{i.i.d.}{independent and identically distributed}
\newacronym{ML}{ML}{maximum likelihood}
\newacronym{PEP}{PEP}{pair-wise error probability}
\newacronym{LLR}{LLR}{log-likelihood ratio}
\newacronym{SNR}{SNR}{signal-to-noise ratio}
\newacronym{SINR}{SINR}{signal-to-interference-plus-noise ratio}

\newacronym{AoI}{AoI}{age of information}
\newacronym{AVP}{AVP}{age-violation probability}
\newacronym{PMF}{PMF}{probability mass function}
\newacronym{CDF}{CDF}{cumulative distribution function}
\newacronym{CCDF}{CCDF}{complementary cumulative distribution function}
\newacronym{SA}{SA}{slotted ALOHA}
\newacronym{IRSA}{IRSA}{irregular repetition slotted ALOHA}
\newacronym{SIC}{SIC}{successive interference cancellation}
\newacronym{PLR}{PLR}{packet loss rate}
\newacronym{DE}{DE}{density evolution}
\newacronym{IoT}{IoT}{Internet of Things}
\newacronym{EH}{EH}{energy harvesting}
\newacronym{CP}{CP}{contention period}

\newacronym{wp}{w.p.}{with probability}
\newacronym{wrt}{w.r.t.}{with respect to}

\newacronym{BEU}{BEU}{best-effort uniform}
\newacronym{TFB}{TFB}{transmit only with full battery}

\IEEEoverridecommandlockouts
\begin{document}
	
	\maketitle
	\begin{abstract} 
		We investigate the age of information (AoI) in a scenario where energy-harvesting devices send status updates to a gateway following the slotted ALOHA protocol and receive no feedback. We let the devices adjust the transmission probabilities based on their current battery level.
		Using a Markovian \revise{approach}, we derive analytically the average AoI. We further provide an approximate analysis for accurate and easy-to-compute approximations of both the average AoI and the age-violation probability (AVP), i.e., the probability that the AoI exceeds a given threshold. We also analyze the average throughput. Via numerical results, we investigate two baseline strategies: transmit a new update whenever possible to exploit every opportunity to reduce the AoI, and transmit
		only when sufficient energy is available to increase the chance of
		successful decoding. The two strategies are beneficial for low and high update-generation rates, respectively. We show that an optimized policy that balances the two strategies  outperforms them significantly in terms of both AoI metrics and throughput. Finally, we show the benefit of decoding multiple packets in a slot using successive interference cancellation and adapting the transmission probability based on both the current battery level and the time elapsed since the last transmission.
	\end{abstract}
	
	\section{Introduction} \label{sec:intro}
	In delay-sensitive \gls{IoT} applications, such as remote sensing, vehicular tracking, and industrial monitoring, devices need to deliver fresh updates about the status of a remote system to a central gateway. To measure the freshness of status updates, the \gls{AoI} metric has been introduced (see, e.g.,~\cite{Kosta2017,Yates2021AoI} and references therein). It captures the time elapsed since the generation of the last update available at 
	the gateway. Early works on the \gls{AoI}, such as~\cite{Kaul2012}, focus on a point-to-point link. More recent extensions to the multi-sensor \gls{IoT} settings consider instead random access protocols based on (slotted) ALOHA~\cite{Berioli2016NOW}, which are widely adopted in commercial applications (e.g., in the satellite
	communication system DVB-RCS2~\cite{ETSI2020DVB} and the low-power wide-area network protocols LoRaWAN~\cite{LoRa} and Sigfox~\cite{SigFox}). The AoI achieved with these protocols has been characterized in, e.g.,~\cite{Yates2017,Yates2020unccordinated,Munari2020modern,Hoang2021AoI,Munari2022_retransmission,Munari2023dynamic}, for the setting where the devices have a stable power supply.
	
	
	In real-world scenarios, IoT devices are often designed for prolonged  low-power operation and sometimes deployed in remote locations, where battery replacement is challenging. A solution to this power supply challenge is energy harvesting, which enables  devices to capture and convert energy from the environment, e.g., thermal, solar, vibration,
	and wireless radio frequency sources, into electrical energy~\cite{Kamalinejad2015wireless,Ku16_energyHarvesting}. Energy harvesting introduces new factors that can significantly affect information freshness, such as the level of energy at the devices when a new update is available, and the time needed for the devices to harvest enough energy. 
	
	Despite its relevance, the impact of energy harvesting on the \gls{AoI} in random-access protocols remains largely unexplored. This paper addresses this gap by characterizing the AoI in a slotted ALOHA system with energy-harvesting devices. 
	
	\subsection{Related Works}
	\subsubsection{\gls{AoI} in Random-Access Networks} 
	The authors of~\cite{Yates2017,Munari2020modern} characterize the average \gls{AoI} of slotted ALOHA over the collision channel. 
	Feedback from the gateway can significantly reduce the \gls{AoI} by allowing each device to adapt its transmission policy to the current \gls{AoI} value~\cite{Yavascan2021,Chen2022AoI,Ahmetoglu2022MiSTA}. The aforementioned works reveal that, for slotted ALOHA, if  devices transmit only if a new update is available, the throughput-maximizing protocol also minimizes the average \gls{AoI}. On the contrary,~\cite{Munari2022_retransmission} shows that, if the devices also retransmit nonfresh updates, 
	the average \gls{AoI} can be improved at the expense of the throughput. The \gls{AoI} of ALOHA has been analyzed for the slot-asynchronous setup in~\cite{Yates2020unccordinated} and the frame-asynchronous setup in~\cite{Wang2023age}. \gls{AoI} analyses have  also been conducted for more advanced ALOHA-based protocols, such as irregular repetition slotted ALOHA~\cite{Munari2020modern,Hoang2021AoI} and frameless ALOHA~\cite{Pan2022CRRA,Munari2023dynamic,Huang2023frameless}. 
	
	\subsubsection{AoI with Energy Harvesting} 
	Most works on the \gls{AoI} with energy harvesting consider a single source sending updates through a point-to-point channel. If the channel is error-free, the \gls{BEU} policy, in which a new update is transmitted whenever the battery is not depleted, minimizes the average \gls{AoI} for the case of infinite battery capacity~\cite{Wu2018optimal}. For finite battery capacity, the optimal online policy has a multi-threshold structure: a new update is transmitted only if the \gls{AoI} exceeds a threshold that decreases monotonically with the available energy~\cite{Arafa2020}. If the channel introduces erasures, the battery capacity is infinite, and there is no feedback,~\cite{Feng2021} shows that the \gls{BEU} policy is still average-\gls{AoI} optimal. On the contrary, periodically retransmitting an update until success is the optimal policy when feedback is available. If the battery has unit capacity and there is no feedback, the optimal policy consists of transmitting a new update if the time elapsed from the last transmission exceeds a threshold that depends on the erasure probability~\cite{Arafa2022}. The aforementioned works assume that the updates are generated at will and transmitted immediately. For the case in which updates are not generated at will,~\cite{Yates2015lazy} and~\cite{Farazi2018} address the setting in which the generated updates are transmitted after a stochastic service delay and according to a service rate, respectively. 
	Beyond the single-source point-to-point channel,~\cite{Arafa2022,Elmagid2022} consider the multi-source setting with a common channel,~\cite{Arafa2019twohop} two-hop networks,~\cite{Chen2021} the two-source multiple access channel,~\cite{Hatami2022} the multi-source error-free channel with request-based updates, and~\cite{Leng2022} 
	the multi-device setting with 
	grant-based 
	protocols. None of these works address the massive random-access setup considered in this paper.
	
	\subsubsection{Random Access with Energy Harvesting} 
	Most existing analyses of ALOHA-based protocols with energy harvesting focus on stability~\cite{Ibrahim2016} and throughput~\cite{Choi2019,Demirhan2019,Haghighat2023,Akyildiz2021}. The authors of~\cite{Bae2017} addressed timely status updates with slotted ALOHA and energy harvesting but analyze the fraction of updates delivered within a deadline rather than the \gls{AoI}. The authors of~\cite{Sleem2020} analyze the \gls{AoI} in a random-access system but consider multiple transmitter-receiver pairs and a different energy harvesting model from our model. 

	\subsection{Contribution}
	We analyze the \gls{AoI} in a scenario where energy-harvesting devices send status updates to a gateway following the slotted ALOHA protocol.\footnote{\revise{We focus on slotted ALOHA because even for this simple protocol, the impact of energy harvesting is not well understood. More advanced protocols, such as feedback-based threshold-ALOHA~\cite{Yavascan2021,Chen2022AoI}, irregular repetition slotted ALOHA~\cite{Liva2011}, and frameless ALOHA~\cite{Stefanovic12} can lead to lower \gls{AoI}.}} There is no feedback from the gateway. We model energy harvesting as an independent Bernoulli process, i.e., a device harvests an energy unit in a slot with a given probability, called the energy harvesting rate. 
	Each device receives readings from a sensor, and thus cannot generate fresh updates at will. Upon receiving a new reading, a device with~$b$ available energy units transmits the update using~$\bt$ energy units \gls{wp} $\pi_{b,\bt}$. 
	The transmitted update is correctly decoded with a probability depending on the transmit power and the level of interference from other devices. 
	Our contributions and main findings are summarized as follows.
	\begin{itemize}[leftmargin=*]
		\item Using a Markovian analysis,\footnote{\revisee{The finite-state Markov chain approach is widely used in {AoI} studies, such as~\cite{Yates2020unccordinated,Munari2020modern,Munari2022_retransmission,Munari2023dynamic,Yavascan2021,Ahmetoglu2022MiSTA,Wang2023age,Wu2018optimal,Yates2015lazy,Farazi2018,Elmagid2022,Hatami2022,Ibrahim2016,Akyildiz2021,Bae2017,Yates2019}, and also well-suited for our problem.}} we 
		derive analytically 
		the average \gls{AoI}; unfortunately, the numerical evaluation of this quantity for parameters of interest is infeasible due to high complexity. 
		
		\item 
		We derive accurate and easy-to-compute approximations 
		of the average \gls{AoI}, the \gls{AVP}, i.e.,  the probability that the \gls{AoI} exceeds a given threshold, as well as the  \gls{PMF} of the \revise{(discretized)} \gls{AoI} at the end of each slot and the peak \gls{AoI}. 
		
		
		\item We conduct numerical experiments where updates are sent over an \gls{AWGN} channel. The receiver either decodes without capture, i.e., decodes only in slots with a single update, or decodes with capture, i.e., decodes in every slot using \gls{SIC}. 
	We consider two baseline strategies: transmit a new update whenever possible (a.k.a. \gls{BEU}~\cite{Wu2018optimal,Feng2021}) to exploit every opportunity to reduce the AoI, and \gls{TFB} to increase the chance of successful decoding. 
	We show that an optimized strategy  significantly outperforms both baselines in terms of the \gls{AoI} metrics and throughput, for both decoding-with-capture and decoding-without-capture cases. \gls{BEU} is close to optimal for low update generation rates but performs poorly for high update generation rates. For the latter scenario, \gls{TFB} is close to optimal for the decoding-without-capture case. However, \gls{TFB} does not benefit from decoding with capture. 
	
	\item  Without capture, 
	the benefit of transmitting with high power vanishes as the power grows large because the successful decoding probability becomes limited by collision. Therefore, the devices should put aside some energy for later transmissions.
	On the contrary, with capture, the devices should transmit with either high or moderate energy, because this facilitates \gls{SIC}. 
	Decoding with capture outperforms decoding without capture for the optimized strategy.
	
	\item The throughput-maximizing strategy entails a loss in the \gls{AoI} metrics, especially for high update generation rates. 
	
	\item 
	A high energy harvesting rate 
	can increase the average \gls{AoI} and \gls{AVP}. In this case, the devices often have enough energy and transmit regardless of the obtainable \gls{AoI} reduction, leading to many transmissions that cause collisions and, even if successful, result in a small \gls{AoI} reduction. 
	This issue can be resolved by progressively increasing the transmission probability 
	after each transmission.
	This prioritizes updates 
	that reduce the \gls{AoI} value considerably if successfully delivered. 
	As shown in~\cite{Yavascan2021,Chen2022AoI} for the case 
	with feedback, adapting the transmission probability to the current \gls{AoI} lowers the average \gls{AoI}. Our results show that adapting the transmission probability to the time elapsed since the last transmission is beneficial also when there is no feedback. 
\end{itemize}

\revise{We extend a previous version of this work~\cite{Hoang2023Globecom} by letting the devices adapt their transmit power rather than always transmitting with all available energy. We also derive the distribution of the peak \gls{AoI} and of the discretized \gls{AoI}. Furthermore, we provide an analysis of the case of always-full battery, for which we propose an adaptive slotted ALOHA protocol without feedback. Different from~\cite{Hoang2023Globecom}, where the devices can only either transmit or harvest energy in a slot, we assume that they can transmit and harvest energy simultaneously. The former setting is relevant if the energy harvesting and radio-frequency transmission functionalities share hardware components, while the latter is relevant if the two functionalities operate independently.} \revisee{This new assumption leads to a minor extension of the Markovian analysis.}

\subsection{Paper Outline and Notation} 
The remainder of the paper is organized as follows. In Section~\ref{sec:model}, we describe the system model and the \gls{AoI} metrics. In Section~\ref{sec:preliminaries}, we present a Markovian analysis of the operation of a device. We then provide an exact and an approximate \gls{AoI} analysis in Sections~\ref{sec:AoI} and~\ref{sec:AoI_approx}, respectively. In Section~\ref{sec:results}, we present numerical results and discussions. We conclude the paper in Section~\ref{sec:conclusions}. \revise{The appendix contains a discussion on the system model, a proof, and some mathematical preliminaries.}

We denote system parameters and constants by uppercase nonitalic letters, e.g.,~$\Ur$, or Greek letters. We denote scalar random variables by uppercase italic letters, e.g.,~$X$, and their realizations by lowercase italic letters, e.g.,~$x$. Vectors are denoted likewise with boldface letters, e.g., a random vector $\Xm$ and its realization~$\xv$. All vectors are column vectors. We use sans-serif, uppercase, and boldface letters, e.g.,~$\Ms$, to denote deterministic matrices. 
By $\Is_m$, $\mathbf{0}_{m}$, and $\mathbf{1}_m$, we denote the $m\times m$ identity matrix, $m\times 1$ all-zero matrix, and $m\times 1$ all-one vector, respectively; the dimension is omitted if it is clear from the context. 
The diagonal matrix with diagonal elements $(d_1,\dots,d_m)$ is denoted by $\diag(d_1,\dots,d_m)$. We denote by $\ind{\cdot}$ the indicator function, $[m:n] = \{m,m+1,\dots,n\}$, 
$[n] \defeq [1:n]$, and $x^+ = \max\{0,x\}$.  
We denote the multinomial distribution with $n$ trials, $k$ events, and event probabilities $\{p_i\}_{i = 1}^k$ by ${\rm Mul}(n,k,\{p_i\}_{i=1}^k)$, and the geometric distribution with success probability $p$ by ${\rm Geo}(p)$. 

\revise{
	\subsection{Reproducible Research} 
	The Matlab code used to evaluate the numerical results is available at: \url{https://github.com/khachoang1412/AoI_slottedALOHA_energyHarvesting}.
}

\section{System Model} \label{sec:model}
We consider a system with $\Ur$ devices attempting to deliver time-stamped status updates (also called packets throughout the paper) to an IoT gateway through a shared channel. Each device receives readings from a sensor, and thus cannot generate fresh updates at will. Updates are generated independently across sensors. Time is slotted and the devices are slot-synchronous. Without loss of generality, we let the slot length be $1$.
Each update transmission spans a slot. A device \revise{may only transmit} if it receives a new sensor reading, which occurs at the beginning of each slot \gls{wp} $\alpha > 0$. 

\subsection{Energy Harvesting}
Each device is equipped with a rechargeable battery with a capacity of $\Er$ energy units. The devices harvest energy from the environment to recharge their batteries. %
As in~\cite{Ibrahim2016,Bae2017,Demirhan2019,Chen2022AoI,Hatami2022}, we model energy harvesting as an independent Bernoulli process. 
In each slot, one energy unit is harvested by a device \gls{wp}~$\EHrate > 0$, independently of the other slots and other devices. 
If the battery is full, the device pauses harvesting. 
We refer to~$\EHrate$ as the energy harvesting rate.
We denote by $\nu_b$ (calculated in Section~\ref{sec:bat_generic_device}) the steady-state probability that the battery level of an arbitrary device is $b \in [0:\Er]$. 

\subsection{Medium Access Protocol}  \label{sec:protocol}
The devices access the medium following the slotted ALOHA protocol. 
Specifically, consider a device with battery level $b \ge 0$. If it has a new update in a slot, it transmits this update using $b_{\rm t} \ge 0$ energy units \gls{wp}~$\pi_{b,b_{\rm t}} \revise{\in [0,1]}$. Otherwise, it stays silent. 
Obviously, $\pi_{b,\bt} = 0$ if $\bt > b$ and $\pi_{b,0} = 1 - \sum_{\bt = 1}^b \pi_{b,\bt}$ is the probability that the device does not transmit despite having a new update. For convenience, we use the convention that $\pi_{b,b_{\rm t}} = 0$ for $\bt < 0$. We denote by $\Pim \revise{\in [0,1]^{\Er \times \Er}}$ the matrix whose $(i,j)$-entry is $\pi_{i,j}$ for $i\in[\Er]$ and $j\in[\Er]$. This lower-triangular matrix contains the design parameters of the protocol. We denote the probability that a device with battery level $b$ transmits using $b_{\rm t}$ energy units by
$\rho_{b,b_{\rm t}} = \alpha \pi_{b,b_{\rm t}} + (1-\alpha)\ind{\bt = 0}.$ When $\bt = 0$, the device stays silent.
We assume that no feedback is provided by the receiver. 

Consider a device that transmits an update with $\bt$ energy units in a slot. We let $L_i\in [\Ur-1]$ be the number of other devices that have battery level $i \in [0:\Er]$, and refer to $\Lm = (L_0, L_1, \dots, L_\Er)$ as the \textit{battery profile} of these devices. 
We denote by $\omega_{\bt,\Lm}$ the probability that the transmitted update is correctly decoded.\footnote{We consider a more general decoding model than the commonly-used collision channel model, where a packet is successfully decoded if there is no interference in the slot and all colliding packets are lost. The collision channel is obtained from our model by setting $\omega_{\bt,\Lm} = \prod_{i=0}^\Er \rho_{i,0}^{L_i}$.} The functional dependency of $\omega_{\bt,\Lm}$ on $(\bt,\Lm)$ captures the impact of the transmit power and  the interference from the other devices. All analytical results in the paper hold for general $\omega_{\bt,\Lm}$. In the numerical experiments  in Section~\ref{sec:results}, we shall instantiate $\omega_{\bt,\Lm}$ by considering a real-valued \gls{AWGN} channel.
At steady state, the average successful-decoding probability of the update is
\begin{equation} \label{eq:avg_w}
	\bar{\omega}_\bt = \E{\omega_{\bt,\Lm}},
\end{equation}
where the expectation is over the steady-state distribution 
${\rm Mul}(\Ur-1,\Er+1,\{\nu_b\}_{b = 0}^\Er)$ of $\Lm$ (see Lemma~\ref{lem:bat_profile} in Section~\ref{sec:Markov_bat_profile}). 
The average throughput, i.e., the average number of packets decoded per slot, is given by
\begin{equation}
	\Tr = \alpha \Ur \sum_{b = 0}^\Er \nu_b \sum_{\bt = 0}^b  \pi_{b,\bt} \bar{\omega}_{\bt}. 
	\label{eq:throughput}
\end{equation}

\subsection{Age of Information} \label{sec:AoI}
\begin{figure}[t!]
	\centering
	\scalebox{.9}{\begin{tikzpicture}[xscale=0.4,yscale=0.3,domain=0:25,samples=400]
				\draw[-latex] (0,0) -- (22,0) node[below] {$t$};
				\draw[-latex] (0,0) -- (0,9.5) node[left,yshift=-.15cm] {$\delta(t)$};
				
				\draw[thick] (2,3.5) -- (5,6.5) -- (5,1) -- (9,5) -- (9,1) -- (17,9) -- (17,1) -- (20,4);
				\draw[dashed] (0,1) -- (21,1);
				\node at (1,3) () {$\dots$};
				\node at (21,3) () {$\dots$};
				\node at (-.5,1.1) () {$1$};
				\node at (-.5,-0.1) () {$0$};
				\node at (-.5,7) () {$\theta$};
				\draw[dashed] (0,7) -- (21,7);
				
				\draw [decorate, decoration = {calligraphic brace}] (17,-.5) -- node[below=.1cm,midway] {$Y$} (9,-.5);
				
				\draw [decorate, decoration = {calligraphic brace}] (17,6.7) -- node[below=.2cm,pos=.8] {$Y\!\!-\!\theta \!+\! 1$} (15,6.7);
			\end{tikzpicture}
		}
	\caption{Example of the AoI process. Here, $Y$ is the time elapsed between two \gls{AoI} refreshes, and $\theta$ is an \gls{AoI} threshold.}
	\label{fig:AoI_process}
\end{figure}
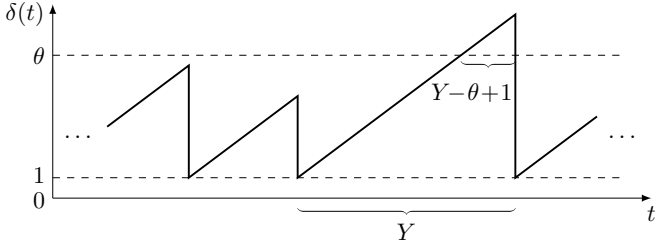

We define the \gls{AoI} of a generic device at time $t$ as
$\delta(t) \defeq t - \tau(t)$,
where $\tau(t)$ is the generation time of the last received update from this device as of time $t$. The corresponding stochastic process is denoted as $\Delta(t)$. 
The \gls{AoI} follows the well-known saw-tooth profile illustrated in~Fig.~\ref{fig:AoI_process}. It grows linearly with time and is reset to $1$ when a new update is successfully decoded. 
Note that $\Delta(t)$ is a continuous-time process that takes values in $\RR$. 
Many \gls{AoI} metrics are defined as a function $F(\Delta) \defeq \lim\limits_{\bar{t} \to \infty}\frac{1}{\bar{t}} \int_{0}^{\bar{t}} f(\Delta(t)) {\rm d}t$ of the process $\Delta(t)$, where $f$ is a given function. For example, the average \gls{AoI} $\bar{\Delta}$ and the \gls{AVP} $\zeta(\theta)$ are defined by setting $f(\Delta(t)) = \Delta(t)$ and $f(\Delta(t)) = \ind{\Delta(t) > \theta}$, respectively. Other classes of \gls{AoI} metrics are obtained by replacing the process $\Delta(t)$ in $F(\Delta)$ by other \gls{AoI}-related processes. Examples include the (discrete) \gls{AoI} value $\widehat\Delta(s)$ at the end of each slot $s$ (before decoding)~\cite{Chen2022AoI}, and the \gls{AoI} value $\widetilde{\Delta}(i)$ just before the $i$th reset of the current \gls{AoI}~\cite{Costa2014}. We refer to $\widehat\Delta(s)$ and $\widetilde{\Delta}(i)$ as the discretized \gls{AoI} and peak \gls{AoI} processes, respectively. 
These two stochastic processes form ergodic Markov chains 
and thus they 
can be characterized via their stationary \gls{PMF}. 


%

In this paper, we use the average \gls{AoI} and \gls{AVP} to assess the \gls{AoI} performance. Furthermore, we also derive the stationary \gls{PMF} of the discretized 
\gls{AoI} and the peak \gls{AoI}. 

\revise{
	We summarize the relevant notation of the paper in Table~\ref{tab:parameters}.}
\begin{table}
	\centering
	\caption{\revise{Descriptions of important symbols}}
	\small
	\def\arraystretch{1.4}
	\revise{
		\begin{tabular}{|p{.7cm} | p{7.2cm} |}
			\hline
			\textbf{$\!\!\!$Symbol$\!\!\!$} & \textbf{Description}  \\
			\hline
			$\Ur$ & number of devices  \\ \hline
			$\Er$ & battery capacity of each device \\\hline
			$\alpha$ & probability that a device has a new update in a slot \\\hline
			$\EHrate$ & probability that a device harvests an energy unit in a slot  \\\hline
			$B$, $b$ & battery level of the device of interest  \\\hline
			$\Lm$, $\ellv$ & battery profile of the remaining $\Ur-1$ devices  \\\hline
			$\pi_{b,\bt}$ & probability that a device transmits with $\bt$ energy units given that it has a new update and has battery level $b$   \\\hline
			$\rho_{b,\bt}$ & probability that a device transmits with $\bt$ energy units given that it has battery level~$b$    \\\hline
			$\omega_{\bt,\Lm}$ & successful decoding probability of an update transmitted with $\bt$ energy units given the battery profile $\Lm$ of the other devices \\\hline
			$\bar{\omega}_\bt$ & average probability that  an update transmitted with $\bt$ energy units is correctly decoded\\\hline
			$\Tr$ & throughput \\\hline
			$\Delta(t)$ & current \gls{AoI} \\\hline
			$\widehat\Delta(s)$ & 
			discretized \gls{AoI}  \\\hline
			$\widetilde\Delta(i)$ & peak \gls{AoI}  \\\hline
			$\bar{\Delta}$ & average \gls{AoI} \\\hline
			$\zeta(\theta)$ & \gls{AVP}
			\\ \hline
		\end{tabular}
	}
	\label{tab:parameters} 
\end{table}

\section{Markov Analysis of the Operation of a Device} \label{sec:preliminaries}


\subsection{Battery Level Evolution of a Generic Device} \label{sec:bat_generic_device}
\begin{figure*}[t!]
	\centering
	\scalebox{.76}{
		\begin{tikzpicture}
			\tikzset{node distance=5.4cm, 
				every state/.style={ 
					semithick,
					fill=gray!10},
				initial text={},     
				double distance=4pt, 
				every edge/.style={  
					draw,
					->,>=stealth',     
					auto,
					semithick}}
     
				\node[state] (b0) {$0$};
				\node[state, right of= b0] (b1) {$1$};
				\node[right of= b1] (mid) {$\dots$};
				\node[state, right of= mid] (b3) {$\!\!\Er\!-\!1\!\!$};
				\node[state, right of= b3] (b4) {$\Er$};
				
				\draw (b0) edge[loop above] node[above,midway] {$1\!-\!\EHrate$} (b0);
				\draw (b1) edge[loop above] node[above,midway] {$\rho_{1,0}(1\!-\!\EHrate) + \rho_{1,1} \EHrate$} (b1);
				\draw (b3) edge[loop above] node[above,midway] {$\rho_{\Er-1,0} (1\!-\!\EHrate) + \rho_{\Er-1,1} \EHrate$} (b3);
				
				\draw (b0) edge[bend left=20] node[above,midway] {$\EHrate$} (b1);
				\draw (b1) edge[bend left=20] node[above,midway] {$\rho_{1,0}\EHrate$} (mid);
				\draw (mid) edge[bend left=20] node[above,midway] {} (b3);
				\draw (b3) edge[bend left=20] node[above,pos=.5] {$\rho_{\Er-1,0} \EHrate$} (b4);
				\draw (b4) edge[loop above] node[above,midway] {$\rho_{\Er,0} + \rho_{\Er,1}\EHrate$} (b4);
				
				\draw (b1) edge[bend left=18] node[above=-.07cm,midway] {$\rho_{1,1} (1-\EHrate)$} (b0);
				\draw (mid) edge[color=OliveGreen,bend left=22] node[above,midway] {} (b0);
				\draw (mid) edge[color=OliveGreen,bend left=18] node[above,midway] {} (b1);
				\draw (b3) edge[color=blue,bend left=25] node[above=-.05cm,pos=.5] {$\rho_{\Er-1,\Er-1}(1\!-\!\EHrate)$} (b0);
				\draw (b3) edge[color=blue,bend left=25] node[above=-.5mm,pos=.5,align=center] {$\rho_{\Er-1,\Er-2}(1\!-\!\EHrate)$ \\ $+ \rho_{\Er-1,\Er-1}\EHrate$} (b1);
				\draw (b3) edge[color=blue,bend left=18]   (mid); 
				\draw (b4) edge[color=red,bend left=27] node[above=-.05cm,pos=.5] {$\rho_{\Er,\Er}(1-\EHrate)$} (b0);
				\draw (b4) edge[color=red,bend left=25] node[above=-.05cm,pos=.45,rotate=3] {$\rho_{\Er,\Er-1}(1\!-\!\EHrate) + \rho_{\Er,\Er}\EHrate$} (b1);
				\draw (b4) edge[color=red,bend left=20]  (mid); 
				\draw (b4) edge[color=red,bend left=18] node[above,pos=.5] {$\rho_{\Er,1}(1\!-\!\EHrate) + \rho_{\Er,2}\EHrate$} (b3);
		\end{tikzpicture}
        }
	\vspace{-.2cm} 
	\caption{Markov chain $M_1$ describing the slot-wise evolution of the battery level of a device.}
	\label{fig:markov_battery}
\end{figure*}
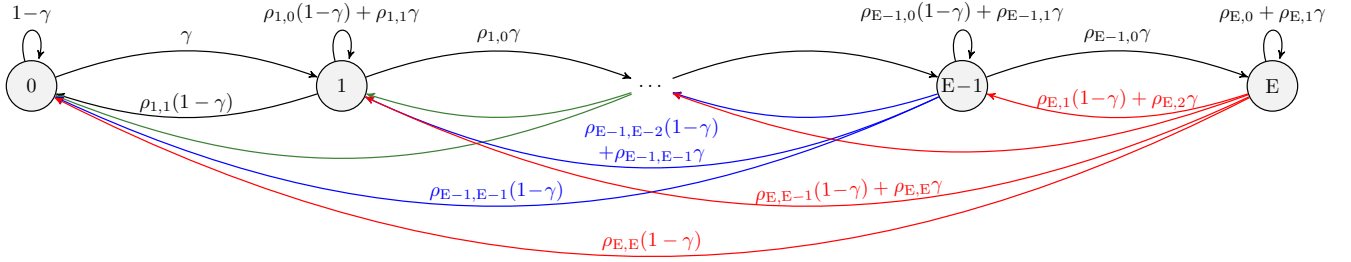

The evolution of the battery level of a generic device is captured by the Markov chain $M_1$ shown in Fig.~\ref{fig:markov_battery}.  Each state represents a battery level. The transition probabilities between the states can be readily computed. Specifically, a device in state $0$ cannot transmit, thus it either remains in this state if it does not harvest energy (\gls{wp} $1-\EHrate$) or moves to state $1$ if an energy unit arrives (\gls{wp} $\EHrate$). A device in state $i \in [\Er]$ moves to state $j \in [0:i+1]$ if it transmits an update with $i-j$ energy units and does not harvest energy (\gls{wp} $\rho_{i,i-j} (1-\EHrate)$), or transmits with $i-j+1$ energy units and harvests one energy unit (\gls{wp} $\rho_{i,i-j+1} \EHrate$). 
If the battery is full, i.e., $i = \Er$, the device remains in state $\Er$ if it does not transmit (\gls{wp} $\rho_{\Er,0}$), or transmits with~$1$ energy unit and harvests one energy unit (\gls{wp} $\rho_{\Er,1}\EHrate$). To summarize, the transition probabilities are given by 
\begin{equation}
	\P{i\to j} = \rho_{i,i-j}(1-\EHrate \ind{(i,j) \ne (\Er,\Er)}) + \rho_{i,i-j+1} \EHrate, \label{eq:p_transition_battery}
\end{equation}
for $ i,j \in [0:\Er]$. Note that $\rho_{0,0} = 1$ and $\rho_{i,k} = 0$ for $k < 0$.
From these transition probabilities, we compute the steady-state distribution $\{\nu_b\}_{b = 0}^\Er$ 
by solving the balance equations\revise{~\cite[Ch.~V]{Kemeny1976_Markov}.}

\subsection{Battery Profile Evolution of the Other $\Ur-1$ Devices} \label{sec:Markov_bat_profile}
The battery profile $\Lm$ of the other devices takes value in
$
\Lc = \big\{(\ell_0,\dots, \ell_\Er) \colon \sum_{i = 0}^\Er \ell_i \!=\! \Ur\!-\!1, \ell_i \in [0:U\!-\!1], i \in [0:\Er]\big\}],
$
with $|\Lc| = \binom{\Ur+\Er-1}{\Er}$.
We now describe the evolution of~$\Lm$ across slots. Let $\ellv' = (\ell'_0, \dots, \ell'_\Er)$ and $\ellv = (\ell_0, \dots, \ell_\Er)$ be the battery profiles at the end of two successive slots. Let also $u_{j,k}$ be the number of devices whose battery level goes from $j$ to $k$. 
We have that 
\begin{align}
	u_{j,k} &\in [0:\min\{\ell'_j,\ell_k\}], \quad  j,k \in [0:\Er],  \label{eq:tmp281} \\
	\ell'_j &= \textstyle\sum_{i=0}^{\min\{j+1,\Er\}} u_{j,i},  \quad j \in [0:\Er],  \\
	\ell_k &= \textstyle\sum_{i=(k-1)^+}^\Er u_{i,k},  \quad k \in [0:\Er]. \label{eq:tmp283} 
\end{align}
The Markov chain $\Lm$ is the composition of $\Ur-1$ independent chains~\cite{Zhou2009composition}, each being identical to $M_1$. We analyze this chain in the next lemma.
\begin{lemma}[Battery profile evolution of $\Ur-1$ devices] \label{lem:bat_profile}
	The transition probabilities of $\Lm$ 
	are given by
	\begin{align} \label{eq:trans_profile}
		\!\!\!\!\P{\ellv' \to \ellv} &= \sum_{\{u_{j,k}\} \colon \text{\eqref{eq:tmp281}--\eqref{eq:tmp283} hold}}  \Bigg(\prod_{j,k \in [0\;:\;\Er]} \P{j\to k}^{u_{j,k}}\Bigg)  \notag\\
		&\quad \cdot \prod_{j=0}^{\Er}\prod_{k=1}^{\min\{j,\Er-1\}} \binom{\ell'_j}{u_{j,0}} \binom{\ell'_j - \sum_{q=0}^{k-1} u_{j,q}}{u_{j,k}},
	\end{align}
	Furthermore, the steady-state distribution of $\Lm$ is ${\rm Mul}(\Ur-1,\Er+1,\{\nu_i\}_{i = 0}^\Er)$.
\end{lemma}
\begin{proof}
	To obtain~\eqref{eq:trans_profile}, we first multiply the probability that $u_{j,k}$ devices go from battery level $j$ to $k$ for $j,k\in[0:\Er]$, 
	with the number of possible partitions of $\ell'_j$ users with battery level~$j$ into sets of $u_{j,0}, u_{j,1}, \dots, u_{j,\min\{j+1,\Er\}}$ users that go from battery level~$j$ to battery levels $0, 1, \dots, \min\{j+1,\Er\}$, respectively, for $j \in [0:\Er]$. 
	We then sum the products over all possible realizations of $\{u_{j,k}\}$. The steady-state distribution of $\Lm$ follows from~\cite[Prop.~1]{Zhou2009composition}.
\end{proof}

\subsection{Markov Chain Describing the Operation of a Generic Device} \label{sec:Markov_G}
We track both the battery level and the \gls{AoI} refresh of a generic device, as well as the battery profile of the other devices. 
We denote by $B\of{s} \in [0:\Er]$ the battery level of the device of interest at the end of slot $s$. We also let $X\of{s} \in \{\refresh,\fail\}$, where $\refresh$ and $\fail$ stand for success and failure, respectively, represent the \gls{AoI} refresh status of the device of interest. Specifically, $X\of{s} = \refresh$ means that the device refreshes its \gls{AoI} value by successfully delivering a new update in slot $s$, and $X\of{s} = \fail$ otherwise. 
Furthermore, we denote the battery profile of the remaining $\Ur-1$ devices at the end of slot $s$ 
by $\Lm\of{s} = (L\of{s}_0, \dots, L\of{s}_\Er) \in \Lc$. 
The Markov chain 
$G\of{s} = (X\of{s},B\of{s},\Lm\of{s})$ fully characterizes the operation of the device across slots. The transition probability from state $(x', b',\ellv')$ to state $(x, b,\ellv)$ of the chain~$G\of{s}$ is given by
\begin{equation} 
	\P{(x', {b'},\ellv') \to (x, b,\ellv)} = \P{x, b'\to b \;|\; \ellv'} \P{\ellv' \to \ellv}. \label{eq:tmp422}
\end{equation}
Here,  $\P{x, b'\!\to\! b \;|\; \ellv'}$ is a shorthand for $\P{X\of{s} \!=\! x, B\of{s} \!=\! b \;|\; B\of{s-1} \!=\! b', \Lm\of{s-1} \!=\! \ellv'}$. It is computed as follows. First, note that $\P{\refresh, b'\!\to\! b \;|\; \ellv'}$  is the probability that, given that the other devices have battery profile $\ellv'$,  the tracked device successfully delivers an update and goes from battery level $b'$ to $b$, i.e., 
\begin{align}
	\P{\refresh, b'\to b \;|\; \ellv'} &=  \omega_{b'-b, \ellv'} \rho_{b',b'-b} (1-\EHrate \ind{(b',b) \ne (\Er,\Er)})  \notag \\
	&\quad +  \omega_{b'-b+1, \ellv'} \rho_{b',b'-b+1}  \EHrate.
\end{align} 
Next, $\P{\fail, b'\to b \;|\; \ellv'}$ is the probability that,  given that the other devices have battery profile $\ellv'$,  the tracked device fails to deliver an update and goes from battery level $b'$ to $b$, i.e., 
\begin{align}
	\P{\fail, b'\to b \;|\; \ellv'} &=   (1 - \omega_{b'-b, \ellv'}) \rho_{b',b'-b}  \notag \\
	&\qquad \cdot (1-\EHrate \ind{(b',b) \ne (\Er,\Er)})  \notag \\
	&\quad + (1 - \omega_{b'-b+1, \ellv'}) \rho_{b',b'-b+1}  \EHrate.
\end{align} 
That is, by accounting for the success/failure probability of update decoding, we replace $\rho_{b',\bt}$ with $\omega_{\bt,\ellv'} \rho_{b',\bt}$ and  $(1-\omega_{\bt,\ellv'})\rho_{b',\bt}$ in $\P{b'\to b}$ given by~\eqref{eq:p_transition_battery} 
to obtain  $\P{\refresh, b'\to b \;|\; \ellv'}$ and $\P{\fail, b'\to b \;|\; \ellv'}$, respectively.
\section{AoI Analysis} \label{sec:AoI}

\subsection{\gls{AoI} Analysis via the Inter-Refresh Time}
We denote by $Y$ the {\em inter-refresh time}, i.e., the number of slots that elapse between two successive \gls{AoI} refreshes for the device of interest (see Fig.~\ref{fig:AoI_process}). 
After a refresh, the current \gls{AoI} is set to $1$ as a packet generated at the start of the current slot is received. 
The average \gls{AoI} 
can be obtained via geometrical arguments as in~\cite[Sec.~II-A]{Yates2021AoI}, \cite[Th.~3]{Yates2019}. The discretized \gls{AoI}, peak \gls{AoI}, and \gls{AVP} have also been studied in different settings in the literature, such as~\cite{Chen2022AoI},~\cite{Costa2014}, and~\cite{Munari2020modern}, respectively. We formally 
express these quantities in terms of $Y$ in the following theorem, and provide a proof in Appendix~\ref{proof:AoI_YZ} for completeness. 
\begin{theorem}[\gls{AoI} metrics in terms of the inter-refresh time distribution] \label{th:AoI_YZ}
	The discretized \gls{AoI} $\widehat\Delta(s)$ and peak \gls{AoI} $\widetilde\Delta(i)$ have stationary \gls{PMF} given by
	\begin{align}
		\P{\widehat\Delta = \delta} &=  \frac{\P{Y > \delta - 2}}{\E{Y}} , \quad \delta = 2,3,\dots \label{eq:PMF_AoI_YZ} \\
		\P{\widetilde{\Delta} = \delta} &= \P{Y = \delta - 1}, \quad \delta = 2,3,\dots \label{eq:PMF_peak_AoI}
	\end{align}
	An \gls{AoI} metric of the form $F(\Delta) = \lim_{\bar{t} \to \infty} \frac{1}{\bar{t}} \int_{0}^{\bar{t}} f(\Delta(t)) {\rm d}t$ can be computed as $F(\Delta) = \frac{1}{\E{Y}} \E{\int_{1}^{Y+1} f(t) {\rm d} t}$. In particular, the average \gls{AoI} $\bar{\Delta}$ and \gls{AVP} $\zeta(\theta)$ are given by
	\begin{align} 
		\bar{\Delta} &= 1 + \frac{\E{Y^2}}{2\E{Y}}, \label{eq:avgAoI_YZ} \\
		\zeta(\theta) &= 1 - \frac{1}{\E{Y}} \bigg(\sum_{y=1}^{\theta-1}y\P{Y=y} + (\theta\!-\!1) \P{Y > \theta-1}\bigg), \notag \\
		&\qquad \theta = 1,2,\dots \label{eq:AVP_YZ} 
	\end{align}
\end{theorem}

\subsection{Average AoI} \label{sec:avg_AoI}
The \gls{AoI} metrics can be derived explicitly. However, as we will explain, their exact computation has high complexity. Due to space limitations, we present only the derivation of the average \gls{AoI} in the following. As implied by~\eqref{eq:avgAoI_YZ}, this entails deriving the moments of $Y$.

\subsubsection{Derivation of $\E{Y}$}
Without loss of generality, we assign index $1$ to the first slot contributing to the current inter-refresh time. We expand $\E{Y}$ as
\begin{align}
	\E{Y} &=\sum_{x \in \{\fail,\refresh\}} \sum_{b \in [0:\Er]} \sum_{\ellv \in \Lc} \E{Y\vert G\of{1} = (x,b,\ellv)} \notag  \\ & \qquad\qquad\qquad\qquad \cdot \mathbb{P}\big[G\of{1} = (x,b,\ellv)\big]. \label{eq:EY_exact}
\end{align}
Noting that the state in a slot with \gls{AoI} refresh is of the form $(\refresh,b',\ellv')$, we have that
\begin{align} 
	&\mathbb{P}\big[G\of{1} = (x,b,\ellv)\big] = \notag \\ 
	&\frac{\sum_{b' \in [0:\Er], \ellv' \in \Lc}\mathbb{P}[(\refresh,b',\ellv') \to (x,b,\ellv)]}{\sum_{g_0 \in \{\refresh, \fail\} \times [0:\Er]\times \Lc} \;\sum_{b' \in [0:\Er], \ellv' \in \Lc}\mathbb{P}[(\refresh,b',\ellv') 
		\to g_0]}. \label{eq:pZ1}
\end{align}
If the \gls{AoI} is refreshed again in slot $1$, i.e., $X\of{1} = \refresh$, 
the inter-refresh time is $1$. Thus, in this case,
\begin{equation}
	\EE\big[Y | G\of{1} = (\refresh,b,\ellv)\big] = 1, \quad \forall b \in [0:\Er], \ellv \in \Lc. \label{eq:tmp385}
\end{equation}
The conditional expectation $\E{Y\vert G\of{1} = (\fail,b,\ellv)}$ can be derived via a first-step analysis~\cite[Sec.~III-4]{TaylKarl98}. Specifically, the inter-refresh time can be computed as the sum of the number of slots until the state $X\of{s}$ becomes $\refresh$. This can be conveniently computed by conditioning on the outcome of the first transition as follows. We first define the probability of having an \gls{AoI} refresh after a state $s$ with $X\of{s} = \fail$, $B\of{s} = b$, and $\Lm\of{s} = \ellv$ as
	$r(b,\ellv) = \sum_{b''\in [0:\Er], \ellv'' \in \Lc} \P{(\fail,b,\ellv) \!\to\! (\refresh,{b''},\ellv'')}$.
We then compute $\E{Y | G\of{1} = (\fail,b,\ellv)}$ as
\begin{align}
	&\E{Y | G\of{1} \!=\! (\fail,b,\ellv)} \notag \\
	&= 1 + \sum_{g\in \{\fail,\refresh\} \times [0:\Er] \times \Lc} \E{Y | G\of{1} \!=\! g} \P{(\fail,b,\ellv) \!\to\! g} \label{eq:tmp390}\\
	&= 1 + r(b,\ellv) + \sum_{b''\in [0:\Er], \ellv'' \in \Lc} \E{Y | G\of{1} = (\fail,{b''},\ellv'')} \notag \\
	&\qquad \qquad \qquad \qquad \qquad  \cdot \P{(\fail,b,\ellv) \to (\fail,{b''},\ellv'')}. 
	\label{eq:tmp393}
\end{align}
In~\eqref{eq:tmp390}, the Markov property ensures that the average duration, once the transition to state $g$ has occurred, is equal to the one that we would have by starting from such state. Let $\ev$ and $\pmb{r}$ be vectors that contain $\E{Y | G\of{1} = (\fail, b,\ellv)}$ and $r(b,\ellv)$, respectively, for all values of $(b,\ellv)$. Let $\Qs$ be a matrix that contains $\P{(\fail,b,\ellv) \to (\fail,{b''},\ellv'')}$ for all $(b,\ellv)$ and $(b'',\ellv'')$. The full-rank system of equations obtained from~\eqref{eq:tmp393} is compactly expressed as $(\Is - \Qs) \ev = \mathbf{1} + \pmb{r}$. Therefore, 
	$\ev = (\Is - \Qs)^{-1} (\mathbf{1} + \pmb{r})$. 
	This result, together with~\eqref{eq:pZ1} and~\eqref{eq:tmp385}, allows us to compute $\E{Y}$ via~\eqref{eq:EY_exact}. 
	
	\subsubsection{Derivation of $\E{Y^2}$}
	$\E{Y^2}$ {can also be} computed via a first-step analysis. Specifically, we observe that
	\begin{equation}
		\E{Y^2 | G\of{1} = (\refresh, b,\ellv)} = 1, \quad \forall b \in [0:E], \ellv \in \Lc, \label{eq:tmp404}
	\end{equation}
	and
	\begin{align}
		&\E{Y^2 | G\of{1} = (\fail, b,\ellv)} \notag \\
		&= 1 + 2 \sum_{g\in \{\fail,\refresh\} \times [0:E] \times \Lc} \E{Y | G\of{1} = g} \P{(\fail, b,\ellv) \!\to\! g}  \notag \\
		&\quad + \sum_{g\in \{\fail,\refresh\} \times [0:E] \times \Lc} \E{Y^2 | Z\of{1} = g} \P{(\fail, b,\ellv) \to g} \notag \\
		&= -1 + 2 \E{Y | Z\of{1} \!=\! (\fail,b,\ellv)} + r(b,\ellv)  \notag \\
		&\quad + \sum_{b''\in [0:E], \ellv'' \in \Lc} \E{Y^2 | Z\of{1} = (\fail, b'',\ellv'')} \notag \\
		&\qquad \qquad \qquad \qquad \cdot  \P{(\fail,b,\ellv) \to (\fail,{b''},\ellv'')}. \label{eq:tmp413}
	\end{align}
	Let now $\ev_2$ be the vector obtained by concatenating the term $\E{Y^2 | Z\of{1} = (\fail, b,\ellv)}$ {for all values of} $(b,\ellv)$. We {can} express~\eqref{eq:tmp413} compactly as $(\Is - \Qs)\ev_2 = -\mathbf{1} + 2\ev+\pmb{r}$. It follows that 
		$\ev_2 = (\Is - \Qs)^{-1} (-\mathbf{1} + 2\ev+\pmb{r})$. 
	Using this,~\eqref{eq:pZ1}, and~\eqref{eq:tmp404}, we compute $\E{Y^2}$ via an expansion analogous to~\eqref{eq:EY_exact}. 
	Finally, {we obtain} the average \gls{AoI} $\bar{\Delta}$ by inserting the computed moments of $Y$ into~\eqref{eq:avgAoI_YZ}.

	\begin{remark}[Complexity Issue]
		The exact computation of $\E{Y}$ and $\E{Y^2}$ requires the evaluation of the transition probabilities between the $n_{\rm s} = 2(\Er+1)\binom{\Ur+\Er-1}{\Er}$ states of the chain $G\of{s}$, and to invert the $(n_{\rm s}/2) \times (n_{\rm s}/2)$ matrix $\Is - \Qs$. These operations become prohibitive for large values of $\Ur$ and $\Er$. 
		This issue motivates us to propose an approximate and low-complexity analysis 
		in the next section.
	\end{remark}
	
	\section{Approximate AoI Analysis} \label{sec:AoI_approx}
	To avoid the complexity issue just highlighted, we ignore the time dependency of the battery profile of the devices whose performance is not tracked. Specifically, we assume the following.
	\begin{simplification} \label{simplification}
		Given a device of interest, the battery profile $\Lm$ of the remaining $\Ur-1$ devices evolves according to a stationary memoryless process across slots. 
	\end{simplification}
	
	This simplification allows us to analyze the behavior of the system for large $\Ur$ and $\Er$ values, and, as we shall see, results in tight approximations of the average AoI and AVP for all scenarios explored in Section~\ref{sec:results}. Under this simplification, the successful-decoding probability of an update transmitted with $\bt$ energy units is given by the average of $\omega_{\bt,\Lm}$ over $\Lm$, i.e., by $\bar{\omega}_\bt$ given in~\eqref{eq:avg_w}.
	This allows us to derive the distribution of $Y$ in closed form, as presented next. 
	
	\subsection{Approximate Distribution of the Inter-Refresh Time $Y$}\label{sec:approxAoI_Y}
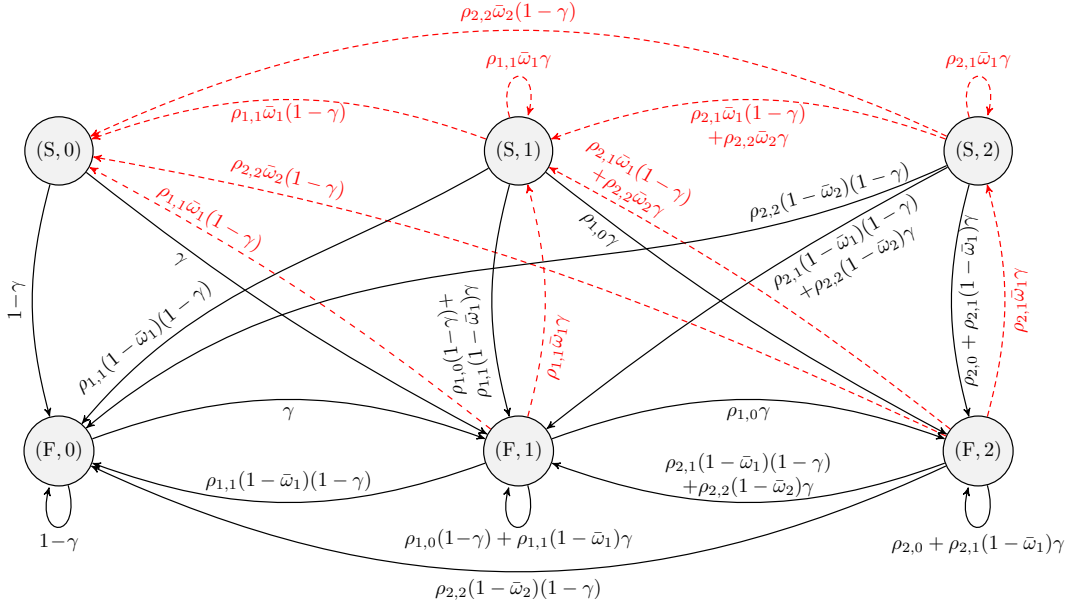
\begin{figure*}[t!]
		\centering
		\scalebox{.76}{
            \begin{tikzpicture}
				\tikzset{node distance=8cm, 
					every state/.style={ 
						semithick,
						fill=gray!10},
					initial text={},     
					double distance=4pt, 
					every edge/.style={  
						draw,
						->,>=stealth',     
						auto,
						semithick}}
    
					\node[state] (F0) {$(\fail,0)$};
					\node[state, right of= F0] (F1) {$(\fail, 1)$};
					\node[state,right of= F1] (F2) {$(\fail,2)$};
					
					\node[state,above=4cm of F0] (S0) {$(\refresh,0)$};
					\node[state, right of= S0] (S1) {$(\refresh, 1)$};
					\node[state, right of= S1] (S2) {$(\refresh,2)$};
					
					\draw (F0) edge[loop below] node[below,midway] {$1\!-\!\EHrate$} (F0);
					\draw (F0) edge[bend left=20] node[below,midway] {$\EHrate$} (F1);
					
					\draw (F1) edge[loop below] node[below=-.1cm,midway] {$\rho_{1,0}(1\!-\!\EHrate) + \rho_{1,1} (1-\bar{\omega}_1) \EHrate$} (F1);
					\draw (F1) edge[bend left=20] node[below,midway] {$\rho_{1,0}\EHrate$} (F2);
					\draw (F1) edge[bend left=20] node[above=.05cm, midway] {$\rho_{1,1} (1-\bar{\omega}_1) (1-\EHrate)$} (F0);
					\draw[color=red,dashed] (F1) edge[bend right=15] node[below,pos=.3,rotate=80] {$\rho_{1,1} \bar{\omega}_1 \EHrate$} (S1);
					\draw[color=red,dashed] (F1) edge[bend right=5] node[above=-.07cm,pos=.75,rotate=-30] {$\rho_{1,1}  \bar{\omega}_1 (1-\EHrate)$} (S0);
					
					\draw (F2) edge[loop below] node[below,midway] {$\rho_{2,0} + \rho_{2,1} (1-\bar{\omega}_1) \EHrate$} (F2);
					\draw (F2) edge[bend left=25] node[below,midway] {$\rho_{2,2}(1-\bar{\omega}_2)(1-\EHrate)$} (F0);
					\draw (F2) edge[bend left=20] node[above=-.05cm,midway,align=center] {$\rho_{2,1}(1-\bar{\omega}_1)(1-\EHrate)$ \\ $+\rho_{2,2}(1-\bar{\omega}_2)\EHrate$} (F1);
					\draw[color=red,densely dashed] (F2) edge[bend right=15] node[below,midway,rotate=90] {$\rho_{2,1} \bar{\omega}_1 \EHrate$} (S2);
					\draw[color=red,densely dashed] (F2) edge[bend right=5] node[above=-.05cm,pos=.85,align=center,rotate=-30] {$\rho_{2,1}\bar{\omega}_1(1-\EHrate)$ \\ $+\rho_{2,2}\bar{\omega}_2\EHrate$} (S1);
					\draw[color=red,densely dashed] (F2) edge[out=155,in=-10] node[above,pos=.8,rotate=-15] {$\rho_{2,2}\bar{\omega}_2 (1-\EHrate)$} (S0);
					
					\draw (S0) edge[bend right=15] node[above,midway,rotate=90] {$1\!-\!\EHrate$} (F0); 
					\draw (S0) edge[bend right=5] node[below,pos=.25,rotate=-40] {$\EHrate$} (F1);
					
					\draw (S1) edge[out=210,in=50] node[above, pos=.8,rotate=38] {$\rho_{1,1} (1-\bar{\omega}_1) (1-\EHrate)$} (F0);
					\draw[color=red,densely dashed] (S1) edge[loop above] node[above,midway] {$\rho_{1,1} \bar{\omega}_1 \EHrate$} (S1);
					\draw (S1) edge[bend right=15] node[above=-.1cm,pos=.65,rotate=97,align=center] {$\rho_{1,0}(1\!-\!\EHrate) +$ \\ $\rho_{1,1} (1-\bar{\omega}_1) \EHrate$} (F1);
					\draw (S1) edge[bend right=5] node[below,pos =.15,rotate=-44] {$\rho_{1,0}\EHrate$} (F2);
					\draw[color=red,densely dashed] (S1) edge[bend right=20] node[below=-.07cm,midway] {$\rho_{1,1}  \bar{\omega}_1 (1-\EHrate)$} (S0);
					
					\draw (S2) edge[out=-155,in=40] node[above,pos=.12,rotate=15] {$\rho_{2,2}(1-\bar{\omega}_2)(1-\EHrate)$} (F0);
					\draw (S2) edge[bend right=5] node[below=-.05cm,pos=.25,align=center,rotate=30] {$\rho_{2,1}(1-\bar{\omega}_1)(1-\EHrate)$ \\ $+\rho_{2,2}(1-\bar{\omega}_2)\EHrate$} (F1);
					\draw (S2) edge[bend right=15] node[below,midway,rotate=90] {$\rho_{2,0} + \rho_{2,1} (1-\bar{\omega}_1) \EHrate$} (F2);
					\draw[color=red,densely dashed] (S2) edge[loop above] node[above,midway] {$\rho_{2,1} \bar{\omega}_1 \EHrate$} (S2);
					\draw[color=red,densely dashed] (S2) edge[bend right=25] node[above,midway] {$\rho_{2,2}\bar{\omega}_2 (1-\EHrate)$} (S0);
					\draw[color=red,densely dashed] (S2) edge[bend right=20] node[below=-.05cm,midway,align=center] {$\rho_{2,1}\bar{\omega}_1(1-\EHrate)$ \\ $+\rho_{2,2}\bar{\omega}_2\EHrate$} (S1);
			\end{tikzpicture}
            }
		\caption{An example of the chain $(X\of{s},B\of{s})$ for $E = 2$. This chain describes the slot-wise evolution of the AoI refresh status and battery level of a device. The transitions that lead to an \gls{AoI} refresh are depicted by red dashed lines.}
		\label{fig:markov_XB}
	\end{figure*}

	Under Simplification~\ref{simplification}, the battery profile $\Lm$ in each slot is drawn independently from the distribution ${\rm Mul}(\Ur-1,\Er+1,\{\nu_b\}_{b = 0}^\Er)$. We therefore only need to track the \gls{AoI} refresh status $X\of{s}$ and battery level $B\of{s}$ of the device of interest. 
	The Markov chain $(X\of{s},B\of{s})$ is obtained from the chain $M_1$ in Fig.~\ref{fig:markov_battery} as follows. We split each battery state $b$ in Fig.~\ref{fig:markov_battery} into two states: \gls{AoI} refresh $(\refresh,b)$ and no \gls{AoI} refresh $(\fail,b)$. Specifically, if the device has battery level $b$ at the end of a slot, the 
	chain moves to state $(\refresh,b)$ if the AoI value is refreshed; otherwise, it moves to state $(\fail,b)$. The transition probabilities between the states 
	can be obtained by accounting for both the battery level transition probabilities $\P{b'\to b}$ and the successful-decoding probability $\bar{\omega}_{\bt}$. Specifically, for $x' \in \{\refresh,\fail\}$, we have that
	\begin{align}
		\P{(x',b') \to (\refresh,b)} &= \bar{\omega}_{b'-b} \rho_{b',b'-b}  (1-\EHrate \ind{(b',b) \!\ne\! (\Er,\Er)})  \notag \\
		&\quad + \bar{\omega}_{b'-b+1} \rho_{b',b'-b+1}  \EHrate, \label{eq:tmp841}\\
		\P{(x',b') \to (\fail,b)} &= (1-\bar{\omega}_{b'-b}) \rho_{b',b'-b}   \notag \\
		&\qquad \cdot  (1-\EHrate \ind{(b',b) \ne (\Er,\Er)})   \notag \\
		&\quad +  (1-\bar{\omega}_{b'-b+1}) \rho_{b',b'-b+1}  \EHrate.\label{eq:tmp842}
	\end{align}
	In Fig.~\ref{fig:markov_XB}, we illustrate 
	$(X\of{s},B\of{s})$ for the case $E=2$.

	We can also obtain the chain $(X\of{s},B\of{s})$ 
	by partitioning the state space of $G\of{s}$ into disjoint subsets of the form $\{(x,b,\ellv) \colon \ellv \in \Lc\}$, and by identifying each subset with a state $(x,b)$ of $(X\of{s}, B\of{s})$. We then compute the transition probabilities 
	as $\P{(x',b') \to (x,b)} = \E{\sum_{\ellv \in \Lc} \P{(x',b',\Lm') \to (x,b,\ellv)}}$, where the expectation is over the steady-state distribution 
	of $\Lm'$. This results in the same formulas as in~\eqref{eq:tmp841} and~\eqref{eq:tmp842}.
	Note that the chain $G\of{s}$ is not \emph{lumpable} with respect to the considered partition.\footnote{A Markov chain $M$ is lumpable with respect to a partition $\{\Ac_i\}_i$ of the states if and only if, for every subset pair $\Ac_i$ and $\Ac_j$, and for every pair of states $m,m'$ in $\Ac_i$, it holds that $\sum_{n \in \Ac_j} \P{m \to n} = \sum_{n \in \Ac_j} \P{m'\to n} = p_{i,j}$~\cite[Th.~6.3.2]{Kemeny1976_Markov}.  The lumped chain with state space $\{\Ac_i\}_i$ and transition probability $\P{\Ac_i \to \Ac_j} = p_{i,j}$ preserves the underlying probabilistic behavior of the original chain $M$. In our case, $G\of{s}$ is not lumpable with respect to the partition $\{(x,b,\ellv) \colon \ellv \in \Lc\}_{x,b}$ because $\sum_{\ellv \in \Lc} \P{(x',b',\ellv') \to (x,b,\ellv)}$ is not constant over $\ellv' \in \Lc$.} However, as we shall see, this partition leads to accurate approximations of the \gls{AoI} metrics. 
	
\begin{figure*}[t!]
		\centering
		\vspace{-.5cm}
		\scalebox{.76}{
			\begin{tikzpicture}
				\tikzset{node distance=8cm, 
					every state/.style={ 
						semithick,
						fill=gray!10},
					initial text={},     
					double distance=4pt, 
					every edge/.style={  
						draw,
						->,>=stealth',     
						auto,
						semithick}}
				
					\node[state] at (0,0) (F0) {$\!(\fail,0)\!$};
					\node[state, right of= F0] (F1) {$\!(\fail, 1)\!$};
					\node[state,right of= F1] (F2) {$\!(\fail,2)\!$};
					
					\node[state,above=4cm of F0] (S0) {$\!(\refresh',0)\!$};
					\node[state, right of= S0] (S1) {$\!(\refresh', 1)\!$};
					\node[state, right of= S1] (S2) {$\!(\refresh',2)\!$};
					
					\node[state] at (21.7,2.3) (R) {$~\refresh''~$};
					
					\draw (F0) edge[loop below] node[below,midway] {$1\!-\!\EHrate$} (F0);
					\draw (F0) edge[bend left=20] node[below,midway] {$\EHrate$} (F1);
					
					\draw (F1) edge[loop below] node[below=-.1cm,midway] {$\rho_{1,0}(1\!-\!\EHrate) + \rho_{1,1} (1-\bar{\omega}_1) \EHrate$} (F1);
					\draw (F1) edge[bend left=20] node[below,midway] {$\rho_{1,0}\EHrate$} (F2);
					\draw (F1) edge[bend left=20] node[above=.05cm, midway] {$\rho_{1,1} (1-\bar{\omega}_1) (1-\EHrate)$} (F0);
					
					\draw (F2) edge[loop below] node[below=-.1cm,midway] {$\rho_{2,0} + \rho_{2,1} (1-\bar{\omega}_1) \EHrate$} (F2);
					\draw (F2) edge[bend left=25] node[below,midway] {$\rho_{2,2}(1-\bar{\omega}_2)(1-\EHrate)$} (F0);
					\draw (F2) edge[bend left=20] node[above=-.05cm,midway,align=center] {$\rho_{2,1}(1-\bar{\omega}_1)(1-\EHrate)$ \\ $+\rho_{2,2}(1-\bar{\omega}_2)\EHrate$} (F1);
					
					\draw (S0) edge[bend right=15] node[above,midway,rotate=90] {$1\!-\!\EHrate$} (F0); 
					\draw (S0) edge[bend right=5] node[below,pos=.25,rotate=-40] {$\EHrate$} (F1);
					
					\draw (S1) edge[out=210,in=50] node[above, pos=.8,rotate=40] {$\rho_{1,1} (1-\bar{\omega}_1) (1-\EHrate)$} (F0);
					\draw (S1) edge[bend right=15] node[above=-.15cm,pos=.65,rotate=97,align=center] {$\rho_{1,0}(1\!-\!\EHrate) +$ \\ $\rho_{1,1} (1-\bar{\omega}_1) \EHrate$} (F1);
					\draw (S1) edge[bend right=5] node[below,pos =.15,rotate=-44] {$\rho_{1,0}\EHrate$} (F2);
					
					\draw (S2) edge[out=-155,in=40] node[above,pos=.12,rotate=15] {$\rho_{2,2}(1-\bar{\omega}_2)(1-\EHrate)$} (F0);
					\draw (S2) edge[bend right=5] node[below=-.05cm,pos=.25,align=center,rotate=30] {$\rho_{2,1}(1-\bar{\omega}_1)(1-\EHrate)$ \\ $+\rho_{2,2}(1-\bar{\omega}_2)\EHrate$} (F1);
					\draw (S2) edge[bend right=15] node[below,midway,rotate=90] {$\rho_{2,0} + \rho_{2,1} (1-\bar{\omega}_1) \EHrate$} (F2);
					
					\draw[color=red, densely dashed] (S2) edge[bend left=10] node[above,midway,rotate=-26] {$\rho_{2,1}\bar{\omega}_1 + \rho_{2,2}\bar{\omega}_2$} (R);
					\draw[color=red, densely dashed] (F2) edge[bend right=10] node[below,midway,rotate=28] {$\rho_{2,1}\bar{\omega}_1 + \rho_{2,2}\bar{\omega}_2$} (R);
					
					\draw[color=red, densely dashed] (S1) edge[out=15, in=115] node[above,pos=.75,rotate=-40] {$\rho_{1,1}\bar{\omega}_1$} (R);
					\draw[color=red, densely dashed] (F1) edge[out=-28, in=-110] node[below,pos=.75,rotate=40] {$\rho_{1,1}\bar{\omega}_1$} (R);
			\end{tikzpicture}
            }
		\caption{An example of the chain $M_2$ for $\Er=2$. This chain describes the slot-wise evolution of the battery level of a device from an \gls{AoI} refresh (state $(\refresh',0)$, $(\refresh',1)$, or $(\refresh',2)$) to the next \gls{AoI} refresh (state $\refresh''$). This chain is obtained from $(X\of{s},B\of{s})$ in Fig.~\ref{fig:markov_XB} by 
			redirecting all transitions that lead to an \gls{AoI} refresh into a new state $\refresh''$.}
		\label{fig:markov_refresh}
	\end{figure*}
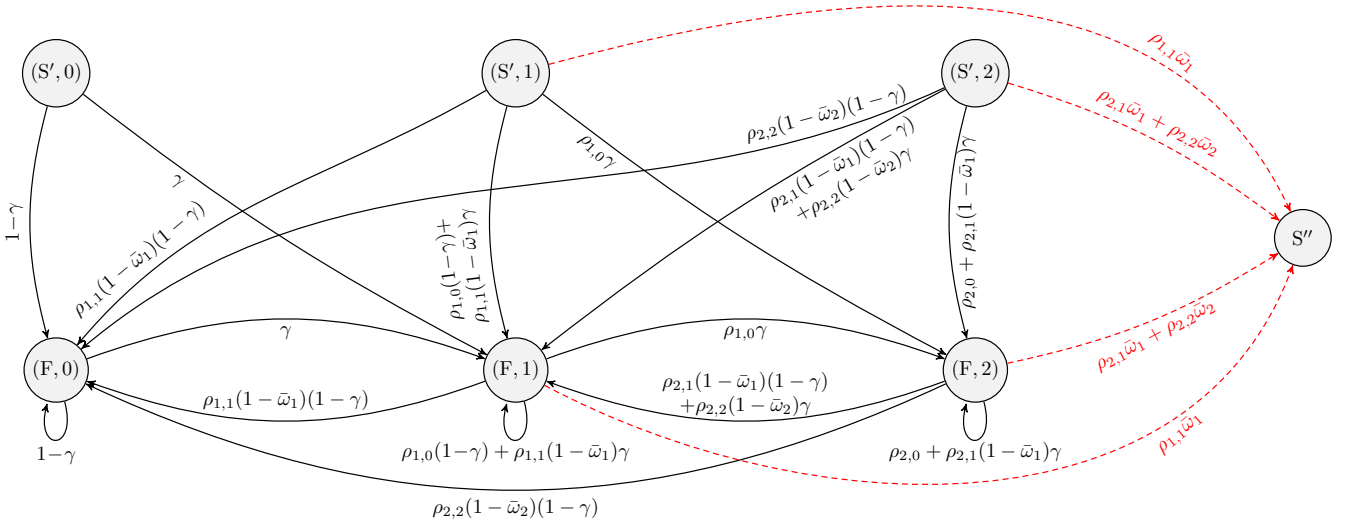

	Next, we find the distribution of $Y$ by analyzing the chain $(X\of{s}, B\of{s})$. To this end, it is convenient to further modify this chain as follows. We split each state $(\refresh,b)$ into two states: $(\refresh',b)$ with only outgoing transitions from $(\refresh,b)$, and $(\refresh'', b)$ with only incoming transitions to $(\refresh,b)$. Furthermore, we combine all states $(\refresh'',b)$, $b \in [0:\Er]$, into a single state~$\refresh''$ that represents an \gls{AoI} refresh. In other words, we redirect all transitions that lead to an \gls{AoI} refresh into a new state $\refresh''$.    
	We refer to the resulting Markov chain as $M_2$, which 
	describes the 
	evolution of the battery level of a device from an \gls{AoI} refresh (i.e., one of the states $(\refresh',b)$, $b \in [0:\Er]$) to the next one (i.e., the state $\refresh''$).
	We depict $M_2$ for the case $\Er = 2$ in Fig.~\ref{fig:markov_refresh}. 

	The chain $M_2$ is a terminating Markov chain (see Appendix~\ref{sec:phase_type}) with an absorbing state $\refresh''$ and $2\Er + 2$ transient states $\big\{(\refresh',0),\dots, (\refresh',{\Er}), (\fail,0),  \dots, (\fail,\Er) \big\}$. We denote the transition probability matrix of $M_2$ as $\bigg[\begin{matrix}
		\Ts ~~ \av \\ \mathbf{0}^\T ~~ 1
	\end{matrix}\bigg]$, where $\Ts$ contains the probabilities of transitions between the transient states and $\av$ contains the probabilities of transitions from the transient states to the absorbing state. We obtain $\Ts$ and $\av$ using~\eqref{eq:tmp841} and~\eqref{eq:tmp842}.
	Observe that $Y$ is the absorption time when starting from state $(\refresh,b)$ \gls{wp} $p_{(\refresh,b)}/\sum_{i=0}^\Er p_{(\refresh,i)}$, $b \in [0:\Er]$, i.e., when the initial probability vector is 
	$\pv_\refresh \defeq \Big(\frac{(p_{(\refresh,0)} \  \dots\  p_{(\refresh,\Er)})}{\sum_{i=0}^\Er p_{(\refresh,i)}}, \mathbf{0}_{\Er +1} \Big)$. Here, $p_{(\refresh,b)}$ is the steady-state probability of the state $(\refresh,b)$ of the chain $(X\of{s}, B\of{s})$. We find $p_{(\refresh,b)}$ by solving the balance equations obtained from the transition probabilities in~\eqref{eq:tmp841} and~\eqref{eq:tmp842}. 
	As a consequence,~$Y$ follows the discrete phase-type distribution characterized in Lemma~\ref{lem:phase_type} in Appendix~\ref{sec:phase_type}. From this, we can readily obtain the \gls{PMF}, \gls{CCDF}, and moments of~$Y$, as presented next.
	\begin{lemma}[Distribution of the inter-refresh time $Y$] \label{lemma:dist_Y}
		Under Simplification~\ref{simplification}, the \gls{PMF} and \gls{CCDF} of 
		$Y$ are given as
		\begin{align} \label{eq:approx_PMF_Y}
			\P{Y = y} &= \pv^\T_\refresh \Ts^{y-1}\av, \quad y = 1,2,\dots \\
			\P{Y > y} &= \pv_\refresh^\T \Ts^y \mathbf{1}_{2\Er+2}, \quad y = 1,2,\dots. \label{eq:approx_CCDF_Y}
		\end{align}
		Furthermore, the first and second moments of $Y$ are given by
		\begin{align} 
			\E{Y} &=  \pv_\refresh^\T (\Is_{2\Er+2}-\Ts)^{-1} \mathbf{1}_{2\Er+2}, \label{eq:approx_EY} \\
			\E{Y^2} &= 2 \pv_\refresh^\T (\Is_{2\Er+2}-\Ts)^{-2}  \mathbf{1}_{2\Er+2} - \E{Y}. \label{eq:approx_EY2}
		\end{align}
	\end{lemma}

	\subsection{Approximate \gls{AoI} Metrics}
	By substituting the \gls{PMF}, \gls{CCDF}, and moments of $Y$ given in Lemma~\ref{lemma:dist_Y} into the expressions of the discretized \gls{AoI} \gls{PMF}, peak \gls{AoI} \gls{PMF}, average \gls{AoI}, and AVP given in Theorem~\ref{th:AoI_YZ}, we obtain readily closed-form approximations for all of these quantities, as shown in the next theorem.
	\begin{theorem}[Approximate AoI metrics] 
		\label{th:approx_AoI}
		Under Simplification~\ref{simplification}, the discretized \gls{AoI} \gls{PMF}, peak \gls{AoI} \gls{PMF}, average \gls{AoI}, and \gls{AVP} are given by
		\begin{align}
			\P{\widehat \Delta = \delta} &= \frac{\pv_\refresh^\T \Ts^{\delta-2}
				\mathbf{1}_{2\Er+2}}{\pv_\refresh^\T(\Is_{2\Er+2}-\Ts)^{-1} \mathbf{1}_{2\Er+2}}, \quad\delta = 2,3,\dots, \label{eq:approx_AoI_PMF} \\
			\P{\widetilde \Delta = \delta} &= \pv^\T_\refresh \Ts^{\delta-2}\av, \quad \delta = 2,3,\dots, \\
			\bar{\Delta} &= \frac{1}{2} +  \frac{\pv_\refresh^\T(\Is_{2\Er+2}-\Ts)^{-2}  \mathbf{1}_{2\Er+2}}{\pv_\refresh^\T(\Is_{2\Er+2}-\Ts)^{-1} \mathbf{1}_{2\Er+2}}, \label{eq:approx_avgAoI} \\
			\zeta(\theta) &= 1 - \frac{\sum_{y=1}^{\theta-1} y  \pv_\refresh^\T\Ts^{y-1}\av + (\theta-1) \pv_\refresh^\T \Ts^{\theta-1} \mathbf{1}_{2\Er+2}}{\pv_\refresh^\T(\Is_{2\Er+2}\!-\!\Ts)^{-1} \mathbf{1}_{2\Er+2}},  \notag \\
			&\qquad  \theta = 1,2\dots
			\label{eq:approx_AVP}
		\end{align}
	\end{theorem}

		\section{The Case of Always-Full Battery} \label{sec:always_full_battery}
		
		We consider the special case where the batteries of the devices are always full. This captures the scenario in which the devices have access to stable energy sources but are subject to a maximum transmit energy constraint $\Er$. This case also captures the regime where $\alpha \ll \EHrate$, which means that the time for a device to fully charge is negligible compared to the time for the device to have a new update.
		
		\subsection{Fixed Transmission Probability Across Slots} \label{sec:fullBat_slotedALOHA}
		For the considered slotted ALOHA protocol, in the case of always-full battery, the design parameter is $\piv = (\pi_{\Er,1}, \dots, \pi_{\Er,\Er})$. For every device, the battery profile of the other devices is fixed to $\ellv_0 = (\mathbf{0}_\Er, \Ur-1)$. An update transmitted with $\bt$ energy units is successfully decoded \gls{wp} $\omega_{\bt,\ellv_0}$. Let ${\omegav} = ({\omega}_{1,\ellv_0}, \dots, {\omega}_{\Er,\ellv_0})$. 
		In a slot, each device successfully delivers a new update \gls{wp} $\xi = \alpha\piv^\T \omegav$, independently across slots. This implies that the inter-refresh time $Y$ follows a geometric distribution with success probability $\xi$. 
		
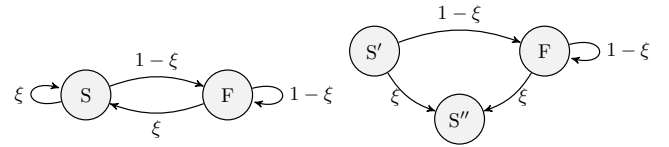
\begin{figure}[t!]
			\centering
			\subcaptionbox{The chain $X\of{s}$ describes the transition between slots with an AoI refresh (state $\refresh$) and slots without one (state~$\fail$). \label{fig:markov_Xs}
					}[.24\textwidth][l]{
						\scalebox{.75}{\begin{tikzpicture}
							\tikzset{node distance=2.5cm, 
								every state/.style={ 
									semithick,
									fill=gray!10},
								initial text={},     
								double distance=4pt, 
								every edge/.style={  
									draw,
									->,>=stealth',     
									auto,
									semithick}}
							{
								\node[state] (S) {$\refresh$};
								\node[state, right of= S] (F) {$\fail$};
								
								\draw (S) edge[loop left] node[left,midway] {$\xi$} (S);
								\draw (F) edge[bend left = 20] node[below,midway] {$\xi$} (S);
								\draw (F) edge[loop right] node[right,midway,align=center] {$1-\xi$} (F);
								
								\draw (S) edge[bend left = 20] node[above,midway] {$1-\xi$} (F);
							}
						\end{tikzpicture}}
					}
					\hfill
					\subcaptionbox{The chain $M_3$ describes the evolution of the device from an AoI refresh (state $S'$) to the next (state~$S''$). \label{fig:markov_M3}
					}[.23\textwidth][c]{
						\scalebox{.75}{\begin{tikzpicture}
							\tikzset{node distance=3cm, 
								every state/.style={ 
									semithick,
									fill=gray!10},
								initial text={},     
								double distance=4pt, 
								every edge/.style={  
									draw,
									->,>=stealth',     
									auto,
									semithick}}
							{
								\node[state] (S) {$\refresh'$};
								\node[state, right of= S] (F) {$\fail$};
								\node[state] at (1.5,-1.2) (R) {$\refresh''$};
								
								\draw (S) edge[bend right=20] node[left,midway] {$\xi$} (R);
								\draw (F) edge[bend left = 20] node[right,midway] {$\xi$} (R);
								\draw (F) edge[loop right] node[right,midway,align=center] {$1-\xi$} (F);
								
								\draw (S) edge[bend left = 20] node[above,midway] {$1-\xi$} (F);
							}
						\end{tikzpicture}}
					}
			\vspace{-.2cm}
			\caption{The Markov chains describing the AoI refresh of a generic device in the case of always-full battery.}
			\label{fig:markov_fullBat}
		\end{figure}

		
		We can also prove that $Y$ follows a geometric distribution using our terminating Markov chain analysis. We start by observing that, in this setup, Simplification~\ref{simplification}  holds. The operation of a generic device is fully characterized by the chain $X\of{s}$ depicted in Fig.~\ref{fig:markov_Xs}. Recall that $X\of{s} = \refresh$  if the device successfully delivers an update in slot $s$, and $X\of{s} = \fail$ otherwise. We further split $\refresh$ into state $\refresh'$ with only outgoing transitions from $\refresh$, and state $\refresh''$ with only incoming transitions to $\refresh$. This yields the chain $M_3$ in Fig.~\ref{fig:markov_M3}. The inter-refresh time~$Y$ is the absorption time of the terminating Markov chain $M_3$. It follows from Lemma~\ref{lem:phase_type} in Appendix~\ref{sec:phase_type} 
		that $\P{Y=y} = \xi (1-\xi)^{y-1}$. 
		
		Substituting the distribution of $Y$ into Theorem~\ref{th:AoI_YZ}, we obtain the closed-form expressions for the discretized \gls{AoI} \gls{PMF}, average \gls{AoI}, and \gls{AVP} stated in the next theorem.
		
		\begin{theorem}[\gls{AoI} metrics with always-full battery] 
			\label{th:approx_AoI_fullBat}
			If the batteries of the devices are always full, the discretized \gls{AoI} \gls{PMF}, peak \gls{AoI} \gls{PMF}, average \gls{AoI}, and \gls{AVP} are given by
			\begin{align}
				\P{\widehat \Delta = \delta} &= \P{\widetilde \Delta = \delta} = \xi(1-\xi)^{\delta-2}, ~ \delta = 2,3,\dots, \label{eq:AoI_PMF_fullBat} \\
				\bar{\Delta} &= \frac{1}{2} +  \frac{1}{\xi} = \frac{1}{2} + \frac{\Ur}{\Tr}, \label{eq:avgAoI_fullBat} \\
				\zeta(\theta) &= (1-\xi)^{\theta-1} = \bigg(1-\frac{\Tr}{\Ur}\bigg)^{\theta-1}, \quad \theta \ge 1. 
				\label{eq:AVP_fullBat}
			\end{align}
		\end{theorem}
		
		Theorem~\ref{th:approx_AoI_fullBat} holds for the general slotted ALOHA setup where a device  delivers a new packet to the gateway in a slot \gls{wp}~$\xi$. The special case of unit-sized battery and collision channel is obtained by setting $\Er=1$ and $\xi = \alpha \pi_1 (1-\alpha\pi_1)^{\Ur-1}$. Therefore, Theorem~\ref{th:approx_AoI_fullBat} generalizes the results for the collision channel reported in~\cite{Munari2020modern,Yates2017}. It follows from Theorem~\ref{th:approx_AoI_fullBat} that the average \gls{AoI},  \gls{AVP}, and throughput are all optimized by the transmission strategy that maximizes~$\xi$. 
		
		%
		

		\subsection{Adaptive Transmission Probability Across Slots} \label{sec:fullBat_thresholdALOHA}
		We now consider a variant of slotted ALOHA where the transmission probability $\piv$ is chosen to be a function of the time elapsed since the last transmission, and of the energy spent in that transmission. We focus on the following multi-threshold scheme. 
		We fix a set of transmission probabilities $ \piv_1, \dots,\piv_\Er$ and let $\piv_0 = \mathbf{0}_\Er$. Note that $\piv = \piv_0$ means that the device stays silent. 
		Assume that the device transmits with $b$ energy units in the current slot. After this slot, the device sets $\piv = \piv_{\Er-b}$, and generates random integers $Q_1, \dots, Q_b$ independently from the ${\rm Geo}(\EHrate)$ distribution. Next, after $Q_i$ slots have elapsed, the device set $\piv$ to $\piv_{\Er-b+i}$, $i\in[b]$, and reset the slot counter. This procedure is repeated after each transmission. We refer to this protocol as {\em multi-threshold slotted ALOHA}. The next theorem 
		relates its \gls{AoI} performance  and the performance of slotted ALOHA with energy harvesting.
		
		\begin{theorem}[Multi-threshold slotted ALOHA] \label{th:thresholdALOHA}
			The \gls{AoI} processes in the two following scenarios are identical: 1) the devices have always-full battery and follow the multi-threshold slotted ALOHA protocol just described; 2) the devices harvest energy with energy harvesting rate $\EHrate$ and follow the slotted ALOHA protocol described in Section~\ref{sec:protocol} with transmission probabilities $\Pim = [\piv_1 \dots \piv_\Er]$. 
		\end{theorem} 
		\begin{proof}
			In energy-harvesting slotted ALOHA, the transmission probability $\piv$ is updated whenever the device changes its battery level. Within a period with no transmission, the battery level changes whenever the device harvests an energy unit. The time duration between changes of $\piv$, when no transmissions occur, is thus identical to the time required for the device to harvest one energy unit. This duration is geometrically distributed with parameter $\EHrate$, i.e., identically distributed to the thresholds in multi-threshold slotted ALOHA with always-full battery. It is easy to verify that the updating rule of $\piv$ in the case of energy harvesting is identical to that in multi-threshold slotted ALOHA. This implies that the statistics of the packet transmission process are the same in the two scenarios. Therefore, the \gls{AoI} processes 
			are identical.
		\end{proof}
		
		For the case of unit-sized battery ($\Er = 1$), multi-threshold slotted ALOHA imposes that each device stays silent after each transmission for a duration of $Q$ slots, and then attempts transmission \gls{wp} $\pi_{1,1}$ whenever it has a new update. This strategy was analyzed for the collision channel with feedback in~\cite{Yavascan2021,Chen2022AoI}, where the backoff $Q$ is set based on the current \gls{AoI} value. 
		There, this strategy was shown to achieve a lower average \gls{AoI} than slotted ALOHA with no backoff. This is because the backoff strategy 
		prioritizes updates that result in a high reduction of \gls{AoI} if successfully delivered.
		In the case of no feedback considered in this paper, however, the devices are not aware of their \gls{AoI}. We therefore let the devices set the backoff based on the time elapsed since their last transmission. 

		Theorem~\ref{th:thresholdALOHA}  implies that 
		the \gls{AoI} metrics of multi-threshold slotted ALOHA with always-full battery can be approximated as in Theorem~\ref{th:approx_AoI}. 
		
		\section{Numerical Experiments and Discussions} \label{sec:results}
		In this section, we assume that the updates are transmitted over a real-valued \gls{AWGN} channel and 
		derive the successful-decoding probability~$\omega_{b,\ellv}$.
		
		\subsection{Channel Model and Successful Delivery Probability} \label{sec:succ_prob}
		We assume that each slot comprises $\Nr$ uses of a real-valued \gls{AWGN} channel. \revise{This channel model is relevant, e.g., in a time-division duplexing system where the gateway broadcasts a downlink pilot signal, each device estimates its channel based on the pilot signal, and active devices pre-equalize their uplink signals based on the channel estimate~\cite{Mei2022,Qiao2024massive}. As in~\cite{Qiao2024massive}, we assume that the channel estimation and pre-equalization steps are perfect, which leads to a Gaussian channel with a known signal-to-noise ratio.} In a slot, active device~$i$ with battery level~$b\of{i}$ transmits with $b\of{i}_{\rm t}$ energy units a signal $\sqrt{\frac{b\of{i}_{{\rm t}}}{\Nr}}\Xm\of{i} \in \RR^\Nr$, with $\|\Xm\of{i}\| = 1$. The received signal in the slot is 
		$
		\Ym = \sum_{i\in \Uc\sub{active}} \sqrt{\frac{b\of{i}_{{\rm t}}}{\Nr}}  \Xm\of{i} + \Zm,
		$
		where $\Uc\sub{active}$ is the set of active devices and
		$\Zm \sim \Nc(\mathbf{0},\sigma^2\Is_\Nr)$ is the \gls{AWGN}. 
		The devices transmit at rate $\Rr$ bit/channel use, i.e., $\Xm\of{i}$ belongs to a codebook containing $2^{\Nr\Rr}$ codewords. We consider shell codes for which the codewords belong to the unit sphere. We analyze two decoding scenarios.
		
		\subsubsection{Without capture} 
		In this scenario, all colliding packets are lost. Decoding is attempted only on packets transmitted in singleton slots. This model allows us to revisit the collision channel model commonly used in modern random-access analyses, e.g.~\cite{Liva2011,Munari2020modern,Demirhan2019}, and to further account for single-user decoding errors due to finite-blocklength effects. Consider an active device that transmits with $\bt$ energy units and assume that the battery profile of the other devices is $\Lm = (L_0, \dots, L_\Er)$. The successful-decoding probability of the device of interest is
		\begin{equation}
			\omega_{\bt,\Lm} = (1-\epsilon_\bt)\textstyle\prod_{i=0}^\Er \rho_{i,0}^{L_i}, \label{eq:w_noCapture}
		\end{equation}
		where $\epsilon_\bt$ is the error probability of decoding the device of interest in a singleton slot. To compute $\epsilon_\bt$, we use that 
		the maximum achievable rate $\Rr^*$ 
		is~\cite[Th.~54]{Polyanskiy2010}
		\begin{equation} \label{eq:rate_FBL}
			\Rr^* = \Cr(\bt) - \sqrt{\frac{\Vr(\bt)}{\Nr}} Q^{-1}(\epsilon_\bt) + O\bigg(\frac{\ln \Nr}{\Nr}\bigg)\,,
		\end{equation}
		where $\Cr(\bt) = \frac{1}{2}\log_2\big(1+\frac{\bt}{\Nr\sigma^2}\big)$, $Q^{-1}(\cdot)$ is the inverse of the Gaussian Q-function $Q(z) = \frac{1}{2\pi}\int_{z}^\infty e^{-t^2/2} {\rm d} t$, and $\Vr(\bt) = \frac{\frac{\bt^2}{\Nr^2\sigma^4} + 2 \frac{\bt}{\Nr\sigma^2}}{2(\bt/(\Nr\sigma^2)+1)^2}\log_2^2(e)$ is the channel dispersion. 
		For a fixed rate~$\Rr$, we use~\eqref{eq:rate_FBL} to approximate $\epsilon_\bt$ as
		$
		\epsilon_\bt \approx Q\Big(\sqrt{\frac{\Nr}{\Vr(\bt)}} (\Cr(\bt) - \Rr) \Big), 
		$
		where we omitted the term $O(\frac{\ln \Nr}{\Nr})$, which is negligible for large $\Nr$. 
		
		
		\subsubsection{With capture} In this case, the receiver attempts to decode every packet transmitted in a slot. 
		Consider an active device that transmits with $\bt$ energy units and let the battery profile of the remaining $\Ur-1$ devices be~$\Lm = (L_0, \dots, L_\Er)$. Furthermore, assume that out of the other devices, $\bar{L}_i$ devices transmit with $i$ energy units. It holds that  $\bar{\Lm} = (\bar{L}_0, \dots, \bar{L}_\Er)$, which we refer to as transmit-energy profile, is the sum of $\Er$ random vectors where the $b$th vector follows a ${\rm Mul}(L_b, \Er+1,\{\rho_{b,i}\}_{i=0}^{\Er})$ distribution, $b \in [\Er]$. 
		Then the interference-to-noise power ratio is $\tilde{\rm P} = \frac{1}{\Nr\sigma^2}\sum_{i=0}^\Er i \bar{L}_i$, and the signal-to-interference-plus-noise ratio is $\bar{\rm P} = \frac{\bt/(\Nr\sigma^2)}{\tilde{\rm P} + 1}$. In this setup, an achievable rate for the device of interest is given as in~\eqref{eq:rate_FBL} with $\Cr(\bt)$ and $\Vr(\bt)$ replaced by $\frac{1}{2}\log_2(1+\bar{\rm P})$ and 
		$\Vr'(\bt,\bar{\Lm}) = \frac{\frac{\bt^2}{\Nr^2\sigma^4}(1 + 2 \tilde{\rm P} 
			+ \tilde{\rm P}^2 - \breve{\rm P}) + 2\frac{\bt}{\Nr\sigma^2}(\tilde{\rm P}+1)^3}{2(\tilde{\rm P}+1)^2(\bt /(\Nr\sigma^2) + \tilde{\rm P} + 1)^2} \log_2^2 e$,
		respectively~\cite[Th.~2]{Scarlett2017}. Here, $\breve{\rm P} = \frac{1}{\Nr^2\sigma^4}\sum_{i=0}^\Er i^2 \bar{L}_i$. Given $\bar{\Lm}$, the error probability of the device can be approximated as
		$
		\epsilon_{\bt,\bar{\Lm}} \approx 
		Q\Big(\sqrt{\frac{\Nr}{V'(\bt,\bar{\Lm})}} \big(\frac{1}{2}\log _2(1\!+\!\bar{\rm P}) \!-\! \Rr\big) \Big). 
		$
		
		We further assume that the receiver employs \gls{SIC}. Specifically, it decodes all devices that transmit with~$\Er$ energy units, removes the decoded packets, then decodes all devices that transmit with $\Er-1$ energy units, and so on. We assume that the decoding of a packet of energy~$j$ is attempted only if all higher-energy packets have been correctly decoded and removed. While decoding the energy-$j$ packet,
		the transmit-energy profile of the interfering devices becomes $\hat{\Lm}\of{j} = [\hat{L}\of{j}_0 \dots \hat{L}\of{j}_\Er]$, where  $\hat{L}\of{\bt}_i = \bar{L}_i \ind{i \le \bt}$ and for $j > \bt$, $\hat{L}\of{j}_i$ takes value $0$ if $i > j$, value $\bar{L}_i - 1$ if $i = j$, value $\bar{L}_i + 1$ if $i = \bt$, and value $\bar{L}_i$ if $i < j, i \le \bt$. 
		It follows that\footnote{To obtain~\eqref{eq:w_capture}, we assume that the events of successfully decoding different packets under  interference from lower-energy packets are independent. This results in an approximation because these events are not independent since the packets are decoded under the same noise.} 
		\begin{equation}
			\omega_{\bt,\Lm} = \mathbb{E}_{\bar{\Lm}}\Big[\big(1 - \epsilon_{\bt,\hat{\Lm}\of{\bt}}\big) \textstyle \prod_{j > \bt}\big(1 - \epsilon_{j,\hat{\Lm}\of{j}}\big)^{\bar{L}_j}\Big]. \label{eq:w_capture}
		\end{equation}
		Note that $\hat{\Lm}\of{\bt}$ is a function of $(\bt,\bar{\Lm})$, and the distribution of $\bar{\Lm}$ is determined by that of~$\Lm$. 
		
		
		In the remainder of this section, we consider a slot length~$\Nr$ of $100$ channel uses, transmission rate~$\Rr$ of $0.8$ bit/channel use, and noise variance $\sigma^2 = -20\;\text{dB}$, unless mentioned otherwise. 
		\revise{The values of $\Nr$ and $\Rr$ are chosen to capture scenarios typical of an \gls{IoT} system~\cite{Durisi16_procIEEE}.}
		We consider two baseline policies: 1) \gls{BEU}, where $\Pim = \Is_{\Er}$, i.e., the devices transmit whenever they have a new update, and 2) \gls{TFB}, where $\Pim = \diag(\mathbf{0}_{\Er-1}, 1)$. In both policies, the devices transmit with all available energy.
		
		\subsection{The Accuracy of the \gls{AoI} Analysis}

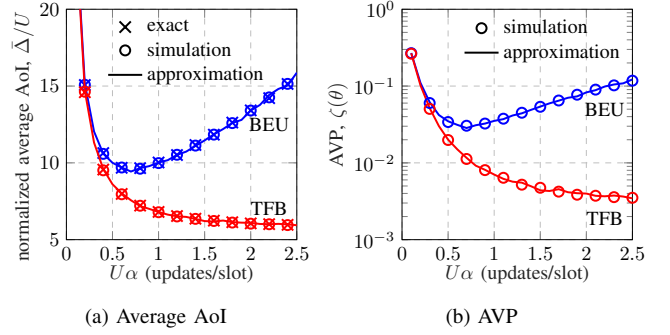
\begin{figure}[t!]
			\centering
			  \hspace{-.2cm}  
\subcaptionbox{Average AoI 
}{
	\begin{tikzpicture}[scale=.75]
		\begin{axis}[%
			width=1.6in,
			height=1.6in,
			clip mode=individual,
			scale only axis,
			unbounded coords=jump,
			xmin=0,
			xmax=2.5,
			xtick= {0, 0.5, 1, 1.5,2,2.5},
			xlabel style={font=\color{white!15!black},yshift=4pt},
			xlabel={$U \alpha$ (updates/slot)},
			ymin=5,
			ymax=20,
			ytick={5,10,15,20},
			yminorticks=true,
			ylabel style={font=\color{white!15!black},yshift=-4pt,xshift=-6pt},
			ylabel={normalized average AoI, $\bar{\Delta}/U$},
			axis background/.style={fill=white},
			title style={font=\bfseries},
			xmajorgrids,
			ymajorgrids,
			legend style={at={(.9,0.99)}, anchor=north east, legend cell align=left, align=left, fill=none,draw=none}
			]
			
			\addplot [line width = 1,color=blue,forget plot]
			table [x index = {0}, y index={1}, col sep=comma]
			{./fig/sim_vs_analytic_capture_pi11.csv}; 
		
		\addplot [line width = 1,color=blue,mark = o, mark color = blue,mark size = 2.5,only marks,forget plot,mark repeat=2]
		table [x index = {0}, y index={2}, col sep=comma]
		{./fig/sim_vs_analytic_capture_pi11.csv}; 
	
	\addplot [line width = 1,color=blue,mark = x, mark color = blue,only marks,forget plot, mark size = 4,mark repeat=2]
	table [x index = {0}, y index={3}, col sep=comma]
	{./fig/sim_vs_analytic_capture_pi11.csv}; 

\addplot [line width = 1,color=red,forget plot]
table [x index = {0}, y index={1}, col sep=comma]
{./fig/sim_vs_analytic_capture_pi01.csv}; 

\addplot [line width = 1,color=red,mark = o, mark color = red,only marks,mark size = 2.5,forget plot,mark repeat=2]
table [x index = {0}, y index={2}, col sep=comma]
{./fig/sim_vs_analytic_capture_pi01.csv}; 

\addplot [line width = 1,color=red,mark = x, mark color = red,only marks,forget plot,mark size = 4,mark repeat=2]
table [x index = {0}, y index={3}, col sep=comma]
{./fig/sim_vs_analytic_capture_pi01.csv}; 

%

\node at (axis cs:2.2,12.5) () {\gls{BEU}}; 
\node at (axis cs:2.2,7) () {\gls{TFB}}; 
\addplot [line width = 1,mark = x, mark color = black,only marks,mark size = 4]
table[row sep=crcr]{%
	-1 0 \\ 
	-1 0 \\ 
};
\addlegendentry{exact};

\addplot [line width = 1,mark = o, mark color = black,mark size = 2.5,only marks]
table[row sep=crcr]{%
-1 0 \\ 
-1 0 \\ 
};
\addlegendentry{simulation};

\addplot [line width = 1,color = black]
table[row sep=crcr]{%
	-1 0 \\ 
	-1 0 \\ 
};
\addlegendentry{approximation};

\end{axis}
\end{tikzpicture}}
\hspace{-.3cm}
\subcaptionbox{\gls{AVP} 
}{
\begin{tikzpicture}[scale=.75]
\begin{axis}[%
width=1.6in,
height=1.6in,
scale only axis,
unbounded coords=jump,
xmin=0,
xmax=2.5,
xtick= {0, 0.5, 1, 1.5,2,2.5},
xlabel style={font=\color{white!15!black},yshift=4pt},
xlabel={$U \alpha$ (updates/slot)},
ymode=log,
ymin=1e-3,
ymax=1,
yminorticks=true,
ylabel style={font=\color{white!15!black},yshift=-4pt},
ylabel={\gls{AVP}, $\zeta(\theta)$},
axis background/.style={fill=white},
title style={font=\bfseries},
xmajorgrids,
ymajorgrids,
legend style={at={(.99,0.99)}, anchor=north east, legend cell align=left, align=left, fill=none,draw=none}
]

\addplot [line width = 1,color=blue,forget plot]
	table [x index = {0}, y index={4}, col sep=comma]
{./fig/sim_vs_analytic_capture_pi11.csv}; 

\addplot [line width = 1,color=blue,mark = o, mark color = blue,only marks,mark size = 2.5,forget plot,mark repeat=2]
	table [x index = {0}, y index={5}, col sep=comma]
{./fig/sim_vs_analytic_capture_pi11.csv}; 

\addplot [line width = 1,color=red,forget plot]
	table [x index = {0}, y index={4}, col sep=comma]
{./fig/sim_vs_analytic_capture_pi01.csv}; 

\addplot [line width = 1,color=red,mark = o, mark color = red,only marks,mark size = 2.5,forget plot,mark repeat=2]
	table [x index = {0}, y index={5}, col sep=comma]
{./fig/sim_vs_analytic_capture_pi01.csv}; 

%

\node at (axis cs:2.2,5e-2) () {\gls{BEU}}; 
\node at (axis cs:2.2,1.9e-3) () {\gls{TFB}}; 

\addplot [line width = 1,mark = o, mark color = black,mark size = 2.5,only marks]
table[row sep=crcr]{%
	-1 1 \\ 
	-1 1 \\ 
};
\addlegendentry{simulation};

\addplot [line width = 1,color = black]
table[row sep=crcr]{%
-1 1 \\ 
-1 1 \\ 
};
\addlegendentry{approximation};
\end{axis}
\end{tikzpicture}}
			\caption{Average AoI and \gls{AVP} vs. average total number of new updates  in a slot ($\Ur\alpha$). Here, $\Ur = 30$, $\EHrate = 0.05$, $\Er = 2$, $\Nr = 100$, $\Rr = 0.8$, $\theta = 1000$, $\sigma^2 = -20$~dB, and the decoding is with capture.
			}
			\label{fig:sim_vs_analytic}
		\end{figure}

		We first verify the accuracy of the exact and approximate analytical AoI analysis by presenting a comparison with simulation results obtained from an implementation of the complete protocol operations over $10^7$ slots. 
		To enable the computation of the exact average \gls{AoI}, we consider a small system with $\Ur = 30$ and $\Er = 2$. We further set $\EHrate = 0.05$ and $\theta = 1000$. 
		In Fig.~\ref{fig:sim_vs_analytic}, we plot the average AoI $\bar{\Delta}$ (normalized by $\Ur$) and \gls{AVP} $\zeta(\theta)$ for the case of decoding with capture and both \gls{BEU} and \gls{TFB} policies. 
		We observe that the approximate average \gls{AoI}~\eqref{eq:approx_avgAoI} matches well both the simulation results and the exact analytical results. The approximate \gls{AVP}~\eqref{eq:approx_AVP} is also in agreement with the simulation results. 

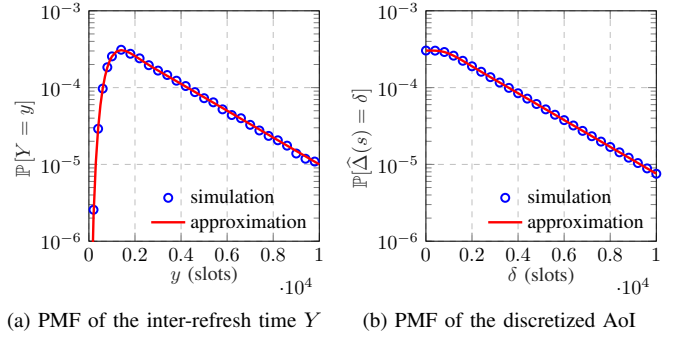
\begin{figure}[t!]
			\centering
			\hspace{-.2cm}
\subcaptionbox{PMF of the inter-refresh time $Y$ 
}{
	\begin{tikzpicture}[scale=.75]
		\begin{axis}[%
			width=1.6in,
			height=1.6in,
			clip mode=individual,
			scale only axis,
			unbounded coords=jump,
			xmin=0,
			xmax=10000,
			xtick= {0, 2000, 4000, 6000, 8000,10000},
			xlabel style={font=\color{white!15!black},yshift=4pt},
			xlabel={$y$ (slots)},
			 ymode=log,
			ymin=1e-6,
			ymax=1e-3,
			yminorticks=true,
			ylabel style={font=\color{white!15!black},yshift=-4pt,xshift=-6pt},
			ylabel={$\P{Y = y}$},
			axis background/.style={fill=white},
			title style={font=\bfseries},
			xmajorgrids,
			ymajorgrids,
			legend style={at={(.99,0.01)}, anchor=south east, legend cell align=left, align=left, fill=none,draw=none}
			]
			
%
			
			\addplot[line width = 1,blue, mark=o, mark options={fill=blue} ,mark size=2pt,only marks, mark repeat=2 
			] 
			table [x index = {0}, y index={1}, col sep=comma]
			{./fig/dist_YZ_capture_noretx_samples.csv}; 
			\addlegendentry{simulation}
			
			\addplot[red, line width=1.2, solid, 
			] 
			table [x index = {0}, y index={2}, col sep=comma]
			{./fig/dist_YZ_capture_noretx_samples.csv}; 
			\addlegendentry{approximation}
		\end{axis}
\end{tikzpicture}}
   \hspace{-.3cm}
\subcaptionbox{PMF of the discretized \gls{AoI} 
}{
	\begin{tikzpicture}[scale=.75]
	\begin{axis}[%
		width=1.6in,
		height=1.6in,
		clip mode=individual,
		scale only axis,
		unbounded coords=jump,
		xmin=0,
		xmax=10000,
		xtick= {0, 2000, 4000, 6000, 8000,10000},
		xlabel style={font=\color{white!15!black},yshift=4pt},
		xlabel={$\delta$ (slots)},
		ymode=log,
		ymin=1e-6,
		ymax=1e-3,
		yminorticks=true,
		ylabel style={font=\color{white!15!black},yshift=-4pt,xshift=-6pt},
		ylabel={$\mathbb{P}[\widehat{\Delta}(s) = \delta]$},
		axis background/.style={fill=white},
		title style={font=\bfseries},
		xmajorgrids,
		ymajorgrids,
		legend style={at={(.99,0.01)}, anchor=south east, legend cell align=left, align=left, fill=none,draw=none}
		]
		
		%
		
		\addplot[line width = 1,blue, mark=o, mark options={fill=blue} ,mark size=2pt,only marks, mark repeat=2, 
		] 
		table [x index = {0}, y index={1}, col sep=comma]
		{./fig/dist_AoI_capture_noretx_samples.csv}; 
		\addlegendentry{simulation}
		
		\addplot[red, line width=1.2, solid, 
		] 
		table [x index = {0}, y index={2}, col sep=comma]
		{./fig/dist_AoI_capture_noretx_samples.csv}; 
		\addlegendentry{approximation}
	\end{axis}
\end{tikzpicture}
}
			\caption{Distribution of the inter-refresh time $Y$ and discretized \gls{AoI} $\widehat{\Delta}(s)$ for $\Ur = 1000$, $\EHrate = 0.005$, $\Er = 8$, $\Nr \!=\! 100$, $\Rr \!=\! 0.8$, $\sigma^2 \!=\! -20$~dB,  $\Pim \!=\! \diag(0,0,0,0,1,1,1,1)$, and decoding with capture. 
			}
			\label{fig:PMF_sim_vs_analytic}
		\end{figure}

		We next consider a larger system with $\Ur = 1000$, $\Er=8$,  and $\EHrate = 0.005$. In Fig.~\ref{fig:PMF_sim_vs_analytic}, we compare the approximate \gls{PMF} of the inter-refresh time $Y$~\eqref{eq:approx_PMF_Y} and the discretized \gls{AoI}~\eqref{eq:PMF_AoI_YZ} with the empirical \gls{PMF} obtained from a simulation 
		over $10^8$ slots for $\Pim = \diag(0,0,0,0,1,1,1,1)$. 
		We see that the approximate \glspl{PMF} match closely the empirical values. 
		
		\revise{In further results reported in Figs.~\ref{fig:metrics_vs_E} and~\ref{fig:metrics_vs_eta}, 
			we observe an excellent agreement between our approximate \gls{AoI} analysis 
			and the simulation results also for decoding without capture and for other sets of parameters.} This confirms that our approximation provides an accurate prediction of the \gls{AoI} metrics. 
		
		\subsection{Optimization of Transmission Probabilities}
		Hereafter, we consider the setting of Fig.~\ref{fig:PMF_sim_vs_analytic}, i.e., $\Ur \!=\! 1000$, $\Er\!=\!8$, $\EHrate \!=\! 0.005$, and we further consider the \gls{AoI} threshold $\theta = 10^4$. 
		We optimize the transmission probabilities $\Pim$ to obtain 
		$\Pim^*_{\bar{\Delta}} = \argmin_{\Pim \in [0,1]^{\Er\times \Er}} \bar{\Delta}$, 
		$\Pim^*_{\zeta} = \argmin_{\Pim \in [0,1]^{\Er\times \Er}} \zeta(\theta),$ 
		and $\Pim^*_{\Tr} = \argmax_{\Pim \in [0,1]^{\Er\times \Er}} \Tr$, where $\Tr$ was defined in~\eqref{eq:throughput}.
		We numerically solve these optimization problems using the Nelder-Mead simplex algorithm~\cite{nelder1965simplex}, a commonly used search method for multidimensional nonlinear optimization. This heuristic method can converge to nonstationary points and is sensitive to the initial values. To circumvent this issue, we run the optimization $200$ times, each with a different initialization.
		
\begin{figure*}[t!]
			\centering
			\input{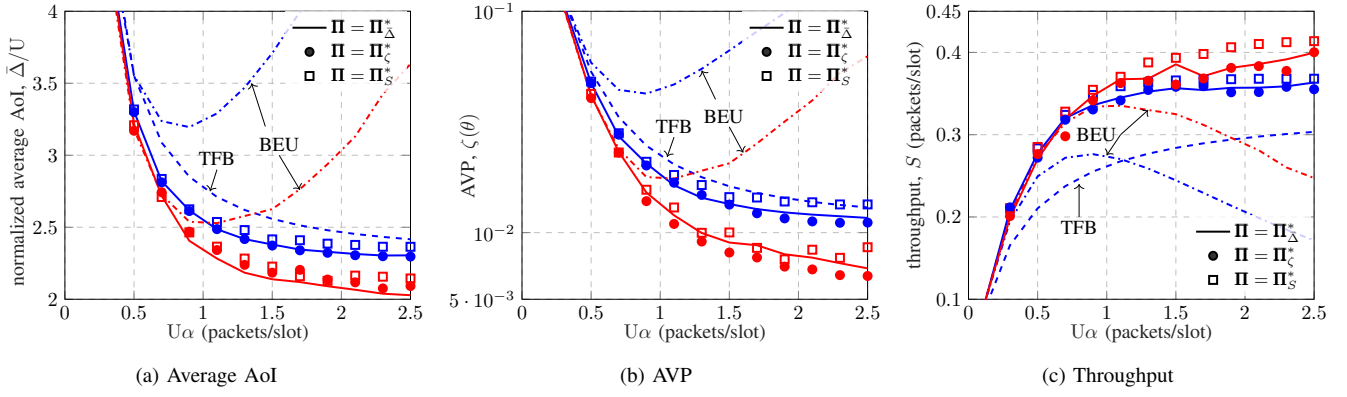}
			\caption{\revise{Approximate} average AoI, \revise{approximate} \gls{AVP}, and throughput vs. average total number of new updates in a slot ($\Ur\alpha$) for different transmission probabilities $\Pim$. 
				Here, $\Ur = 1000$, $\EHrate = 0.005$, $\Er = 8$, $\Nr = 100$, $\Rr = 0.8$, $\theta = 10^4$, and $\sigma^2 = -20$~dB. The red and blue curves represent decoding with and without capture, respectively.}
			\label{fig:opt_pi_eta.005}
		\end{figure*}

		In Fig.~\ref{fig:opt_pi_eta.005}, we plot the minimized \revise{approximate} average \gls{AoI}, minimized \revise{approximate} \gls{AVP}, and maximized throughput as functions of $\Ur\alpha$, and compare them with the corresponding metrics achieved by the \gls{BEU} and \gls{TFB} policies. We consider both decoding with and without capture.\footnote{In the considered setting, for \gls{TFB}, the performance with capture coincides with that without capture. Indeed, since all devices transmit with high power, decoding 
			under interference from other devices fails with high probability.} 
		The optimized $\Pim$ leads to significant improvement in all three metrics. The \gls{BEU} strategy is close to optimal when $\Ur\alpha$ is small, especially with capture. However, it becomes highly suboptimal when $\Ur\alpha$ increases since it causes many collisions. In contrast, the \gls{TFB} policy performs closely to the optimal policy (without capture) for large $\Ur\alpha$. With capture, the minimized average AoI and maximized throughput improve by about $11.7\%$ and $12.4\%$, respectively, for $\Ur\alpha = 2.5$, compared to decoding without capture.
		
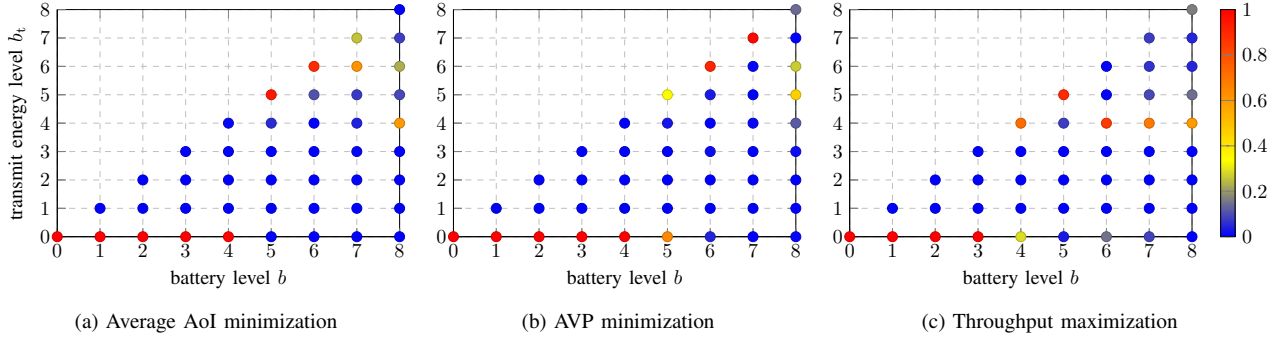
\begin{figure*}[t!]
			\centering
			\subcaptionbox{Average AoI minimization 
}{\begin{tikzpicture}[scale=.75]
		\begin{axis}[grid=major,view={0}{90},
			width=3in,
			height=2.2in,
			domain = 0:8,
			xtick={0,1,2,3,4,5,6,7,8},
			ytick={0,1,2,3,4,5,6,7,8},
			y domain = 1:8,
			y domain = 1:8,
			xlabel={battery level $b$},
			ylabel={transmit energy level $\bt$}, 
			]
			\addplot3+[only marks,scatter,mark size = 2.5] 
			table[row sep=\\]
			{%
				x y z \\
				0 0 1.00000000e+00 \\ 
				1 0 9.99969634e-01 \\ 
				1 1 3.03663133e-05 \\ 
				2 0 9.99944301e-01 \\ 
				2 1 2.12365882e-08 \\ 
				2 2 5.56774206e-05 \\ 
				3 0 9.96923071e-01 \\ 
				3 1 3.88463528e-05 \\ 
				3 2 2.94397798e-05 \\ 
				3 3 3.00864292e-03 \\ 
				4 0 9.90909705e-01 \\ 
				4 1 6.69617202e-05 \\ 
				4 2 1.10765474e-05 \\ 
				4 3 8.13100769e-04 \\ 
				4 4 8.19915631e-03 \\ 
				5 0 1.97983366e-03 \\ 
				5 1 1.83218280e-05 \\ 
				5 2 2.90030260e-04 \\ 
				5 3 1.94899007e-04 \\ 
				5 4 5.58023835e-02 \\ 
				5 5 9.41714532e-01 \\ 
				6 0 3.47251452e-03 \\ 
				6 1 1.21458222e-04 \\ 
				6 2 3.61433492e-05 \\ 
				6 3 4.65470357e-06 \\ 
				6 4 2.89973404e-03 \\ 
				6 5 1.07348501e-01 \\ 
				6 6 8.86116994e-01 \\ 
				7 0 2.79079923e-03 \\ 
				7 1 1.18752547e-05 \\ 
				7 2 4.47676362e-03 \\ 
				7 3 6.98942526e-03 \\ 
				7 4 4.20830968e-02 \\ 
				7 5 7.49458723e-02 \\ 
				7 6 6.15010060e-01 \\ 
				7 7 2.53692108e-01 \\ 
				8 0 1.15854644e-04 \\ 
				8 1 4.58320755e-04 \\ 
				8 2 1.56784184e-03 \\ 
				8 3 1.09013530e-03 \\ 
				8 4 5.96847866e-01 \\ 
				8 5 9.33697710e-02 \\ 
				8 6 2.26625974e-01 \\ 
				8 7 7.99003236e-02 \\ 
				8 8 2.39130846e-05 \\ 
			};
		\end{axis}
	\end{tikzpicture}
}
\subcaptionbox{AVP minimization 
}{\begin{tikzpicture}[scale=.75]
		\begin{axis}[grid=major,view={0}{90},
			width=3in,
			height=2.2in,
			domain = 0:8,
			xtick={0,1,2,3,4,5,6,7,8},
			ytick={0,1,2,3,4,5,6,7,8},
			y domain = 1:8,
			y domain = 1:8,
			xlabel={battery level $b$},
			]
			\addplot3+[only marks,scatter,mark size = 2.5] 
			table[row sep=\\]
			{%
				x y z \\
				0 0 1.00000000e+00 \\ 
				1 0 9.99785426e-01 \\ 
				1 1 2.14573901e-04 \\ 
				2 0 9.98309877e-01 \\ 
				2 1 2.49198994e-04 \\ 
				2 2 1.44092373e-03 \\ 
				3 0 9.99245241e-01 \\ 
				3 1 2.02749509e-05 \\ 
				3 2 1.66891829e-08 \\ 
				3 3 7.34467447e-04 \\ 
				4 0 9.93228720e-01 \\ 
				4 1 2.15913187e-04 \\ 
				4 2 1.92193854e-06 \\ 
				4 3 5.15537359e-03 \\ 
				4 4 1.39807156e-03 \\ 
				5 0 6.43366231e-01 \\ 
				5 1 8.46399264e-05 \\ 
				5 2 1.81817868e-04 \\ 
				5 3 9.35711686e-06 \\ 
				5 4 2.88997497e-02 \\ 
				5 5 3.27458204e-01 \\ 
				6 0 4.75726812e-02 \\ 
				6 1 2.75185180e-04 \\ 
				6 2 5.53046771e-05 \\ 
				6 3 1.48168559e-02 \\ 
				6 4 9.02946428e-04 \\ 
				6 5 3.39591858e-02 \\ 
				6 6 9.02417841e-01 \\ 
				7 0 7.41595365e-05 \\ 
				7 1 7.19514314e-04 \\ 
				7 2 6.34144251e-04 \\ 
				7 3 6.56454716e-04 \\ 
				7 4 8.21134407e-03 \\ 
				7 5 1.38977774e-04 \\ 
				7 6 6.08680945e-03 \\ 
				7 7 9.83478596e-01 \\ 
				8 0 6.59406259e-03 \\ 
				8 1 8.85227293e-05 \\ 
				8 2 2.14012064e-04 \\ 
				8 3 1.66128142e-02 \\ 
				8 4 1.16776808e-01 \\ 
				8 5 4.52269161e-01 \\ 
				8 6 2.65476136e-01 \\ 
				8 7 3.95230185e-03 \\ 
				8 8 1.38016181e-01 \\ 
			};
		\end{axis}
	\end{tikzpicture}
}
\subcaptionbox{Throughput maximization 
}{\begin{tikzpicture}[scale=.75]
		\begin{axis}[grid=major,view={0}{90},
			width=3in,
			height=2.2in,
			domain = 0:8,
			xtick={0,1,2,3,4,5,6,7,8},
			ytick={0,1,2,3,4,5,6,7,8},
			y domain = 1:8,
			y domain = 1:8,
			xlabel={battery level $b$},
										colorbar,
					colorbar style={width=.1in}
			]
			\addplot3+[only marks,scatter,mark size = 2.5] 
			table[row sep=\\]
			{%
				x y z \\
				 0 0 1.00000000e+00 \\ 
				1 0 9.99999937e-01 \\ 
				1 1 6.31062175e-08 \\ 
				2 0 9.99999890e-01 \\ 
				2 1 8.84149820e-09 \\ 
				2 2 1.01297277e-07 \\ 
				3 0 9.99999846e-01 \\ 
				3 1 6.30912100e-09 \\ 
				3 2 7.20643756e-08 \\ 
				3 3 7.54441623e-08 \\ 
				4 0 2.86433001e-01 \\ 
				4 1 6.39552145e-09 \\ 
				4 2 3.66298608e-08 \\ 
				4 3 3.04624609e-06 \\ 
				4 4 7.13563910e-01 \\ 
				5 0 3.08065876e-02 \\ 
				5 1 7.05577065e-08 \\ 
				5 2 2.80202909e-07 \\ 
				5 3 5.23245175e-06 \\ 
				5 4 7.99915631e-02 \\ 
				5 5 8.89196266e-01 \\ 
				6 0 1.49118904e-01 \\ 
				6 1 1.03344857e-07 \\ 
				6 2 6.38163763e-07 \\ 
				6 3 1.34212018e-06 \\ 
				6 4 8.46701812e-01 \\ 
				6 5 2.99411038e-03 \\ 
				6 6 1.18309048e-03 \\ 
				7 0 8.58052656e-02 \\ 
				7 1 9.68731693e-08 \\ 
				7 2 1.58330018e-07 \\ 
				7 3 4.62815146e-05 \\ 
				7 4 6.77274604e-01 \\ 
				7 5 9.44134903e-02 \\ 
				7 6 6.44746220e-02 \\ 
				7 7 7.79854813e-02 \\ 
				8 0 1.64354119e-04 \\ 
				8 1 1.06095744e-06 \\ 
				8 2 1.80725703e-08 \\ 
				8 3 1.96335173e-05 \\ 
				8 4 5.93316903e-01 \\ 
				8 5 1.43929147e-01 \\ 
				8 6 5.00020277e-02 \\ 
				8 7 4.79009122e-02 \\ 
				8 8 1.64665944e-01 \\ 
			};
		\end{axis}
	\end{tikzpicture}
}
			\caption{The probability $\pi_{b,\bt}$ (represented by colors) that a device transmits with $\bt$ energy units if it has battery level $b$ and has a new update for the strategies that minimize the average AoI, minimize the \gls{AVP}, and maximize the throughput for the setting in Fig.~\ref{fig:opt_pi_eta.005} without capture and with $\alpha \Ur = 2.5$.}
			\label{fig:opt_pi}
		\end{figure*}

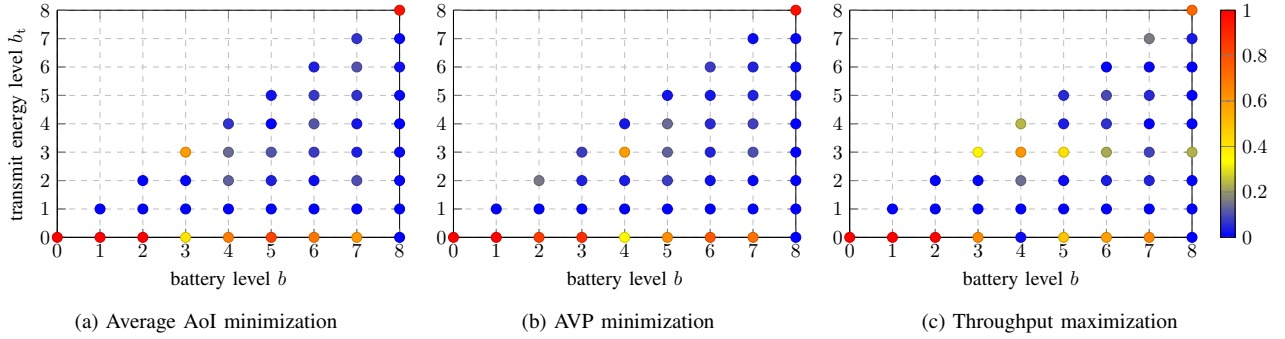
\begin{figure*}[t!]
			\centering
			\subcaptionbox{Average AoI minimization 
}{\begin{tikzpicture}[scale=.75]
		\begin{axis}[grid=major,view={0}{90},
			width=3in,
			height=2.2in,
			domain = 0:8,
			xtick={0,1,2,3,4,5,6,7,8},
			ytick={0,1,2,3,4,5,6,7,8},
			y domain = 1:8,
			y domain = 1:8,
			xlabel={battery level $b$},
			ylabel={transmit energy level $\bt$}, 
			]
			\addplot3+[only marks,scatter,mark size = 2.5] 
			table[row sep=\\]
			{%
				x y z \\
					 0 0 1.00000000e+00 \\ 
					1 0 1.00000000e+00 \\ 
					1 1 0.00000000e+00 \\ 
					2 0 9.99853945e-01 \\ 
					2 1 0.00000000e+00 \\ 
					2 2 1.46055373e-04 \\ 
					3 0 4.00807285e-01 \\ 
					3 1 0.00000000e+00 \\ 
					3 2 7.68515696e-03 \\ 
					3 3 5.91507558e-01 \\ 
					4 0 6.68804337e-01 \\ 
					4 1 0.00000000e+00 \\ 
					4 2 1.25672594e-01 \\ 
					4 3 1.39872755e-01 \\ 
					4 4 6.56503136e-02 \\ 
					5 0 8.26558119e-01 \\ 
					5 1 0.00000000e+00 \\ 
					5 2 3.88280441e-02 \\ 
					5 3 1.09643071e-01 \\ 
					5 4 2.74775336e-03 \\ 
					5 5 2.22230130e-02 \\ 
					6 0 6.88136671e-01 \\ 
					6 1 0.00000000e+00 \\ 
					6 2 1.54653621e-02 \\ 
					6 3 7.50348954e-02 \\ 
					6 4 1.11883851e-01 \\ 
					6 5 7.23713492e-02 \\ 
					6 6 3.71078719e-02 \\ 
					7 0 5.97227434e-01 \\ 
					7 1 0.00000000e+00 \\ 
					7 2 1.01192591e-01 \\ 
					7 3 3.86800959e-02 \\ 
					7 4 2.63318125e-02 \\ 
					7 5 6.34984494e-02 \\ 
					7 6 1.07594722e-01 \\ 
					7 7 6.54748950e-02 \\ 
					8 0 0.00000000e+00 \\ 
					8 1 0.00000000e+00 \\ 
					8 2 2.00350587e-03 \\ 
					8 3 8.32891124e-03 \\ 
					8 4 9.35360165e-05 \\ 
					8 5 7.43924583e-03 \\ 
					8 6 1.93328077e-02 \\ 
					8 7 7.13366242e-03 \\ 
					8 8 9.55668331e-01 \\ 
			};
		\end{axis}
	\end{tikzpicture}
}
\subcaptionbox{AVP minimization 
}{\begin{tikzpicture}[scale=.75]
		\begin{axis}[grid=major,view={0}{90},
			width=3in,
			height=2.2in,
			domain = 0:8,
			xtick={0,1,2,3,4,5,6,7,8},
			ytick={0,1,2,3,4,5,6,7,8},
			y domain = 1:8,
			y domain = 1:8,
			xlabel={battery level $b$},
			]
			\addplot3+[only marks,scatter,mark size = 2.5] 
			table[row sep=\\]
			{%
				x y z \\
				 0 0 1.00000000e+00 \\ 
				1 0 1.00000000e+00 \\ 
				1 1 0.00000000e+00 \\ 
				2 0 8.44079527e-01 \\ 
				2 1 0.00000000e+00 \\ 
				2 2 1.55920473e-01 \\ 
				3 0 8.82680020e-01 \\ 
				3 1 0.00000000e+00 \\ 
				3 2 3.93285572e-02 \\ 
				3 3 7.79914228e-02 \\ 
				4 0 3.45249903e-01 \\ 
				4 1 0.00000000e+00 \\ 
				4 2 3.50594896e-02 \\ 
				4 3 5.90991970e-01 \\ 
				4 4 2.86986373e-02 \\ 
				5 0 6.26864409e-01 \\ 
				5 1 0.00000000e+00 \\ 
				5 2 7.86419802e-02 \\ 
				5 3 1.29086173e-01 \\ 
				5 4 1.39317942e-01 \\ 
				5 5 2.60894957e-02 \\ 
				6 0 7.74639585e-01 \\ 
				6 1 0.00000000e+00 \\ 
				6 2 1.99209819e-02 \\ 
				6 3 4.54064603e-02 \\ 
				6 4 5.51511864e-02 \\ 
				6 5 2.76498094e-02 \\ 
				6 6 7.72319770e-02 \\ 
				7 0 7.08605268e-01 \\ 
				7 1 0.00000000e+00 \\ 
				7 2 7.60035274e-03 \\ 
				7 3 1.03338572e-01 \\ 
				7 4 8.41260312e-02 \\ 
				7 5 4.13185909e-02 \\ 
				7 6 4.25135245e-02 \\ 
				7 7 1.24976604e-02 \\ 
				8 0 0.00000000e+00 \\ 
				8 1 0.00000000e+00 \\ 
				8 2 1.64195861e-03 \\ 
				8 3 1.26319651e-03 \\ 
				8 4 2.42662558e-03 \\ 
				8 5 1.26123935e-02 \\ 
				8 6 9.48521931e-03 \\ 
				8 7 1.10098514e-02 \\ 
				8 8 9.61560755e-01 \\ 
			};
		\end{axis}
	\end{tikzpicture}
}
\subcaptionbox{Throughput maximization 
}{\begin{tikzpicture}[scale=.75]
		\begin{axis}[grid=major,view={0}{90},
			width=3in,
			height=2.2in,
			domain = 0:8,
			xtick={0,1,2,3,4,5,6,7,8},
			ytick={0,1,2,3,4,5,6,7,8},
			y domain = 1:8,
			y domain = 1:8,
			xlabel={battery level $b$},
										colorbar,
					colorbar style={width=.1in}
			]
			\addplot3+[only marks,scatter,mark size = 2.5] 
			table[row sep=\\]
			{%
				x y z \\
				0 0 1.00000000e+00 \\ 
				1 0 1.00000000e+00 \\ 
				1 1 0.00000000e+00 \\ 
				2 0 9.96659879e-01 \\ 
				2 1 0.00000000e+00 \\ 
				2 2 3.34012139e-03 \\ 
				3 0 6.35628113e-01 \\ 
				3 1 0.00000000e+00 \\ 
				3 2 5.48156705e-03 \\ 
				3 3 3.58890320e-01 \\ 
				4 0 1.29600066e-02 \\ 
				4 1 0.00000000e+00 \\ 
				4 2 1.44146267e-01 \\ 
				4 3 6.05559100e-01 \\ 
				4 4 2.37334627e-01 \\ 
				5 0 4.85238877e-01 \\ 
				5 1 0.00000000e+00 \\ 
				5 2 1.44903507e-02 \\ 
				5 3 4.09529102e-01 \\ 
				5 4 3.66183831e-02 \\ 
				5 5 5.41232876e-02 \\ 
				6 0 5.83641308e-01 \\ 
				6 1 0.00000000e+00 \\ 
				6 2 3.76182791e-02 \\ 
				6 3 2.21245225e-01 \\ 
				6 4 6.08230017e-02 \\ 
				6 5 9.60098520e-02 \\ 
				6 6 6.62334309e-04 \\ 
				7 0 6.46161240e-01 \\ 
				7 1 0.00000000e+00 \\ 
				7 2 4.27682159e-02 \\ 
				7 3 8.62833486e-02 \\ 
				7 4 2.09087540e-03 \\ 
				7 5 5.28619391e-02 \\ 
				7 6 9.02450949e-03 \\ 
				7 7 1.60809872e-01 \\ 
				8 0 0.00000000e+00 \\ 
				8 1 0.00000000e+00 \\ 
				8 2 4.29045839e-04 \\ 
				8 3 2.32732551e-01 \\ 
				8 4 3.30284339e-03 \\ 
				8 5 1.14188239e-03 \\ 
				8 6 3.72183926e-03 \\ 
				8 7 3.04775995e-02 \\ 
				8 8 7.28194238e-01 \\ 
			};
		\end{axis}
	\end{tikzpicture}
}
			\caption{Same as Fig.~\ref{fig:opt_pi} but with capture.}
			\label{fig:opt_pi_capture}
		\end{figure*}

		
		In Figs.~\ref{fig:opt_pi} and~\ref{fig:opt_pi_capture}, we 	present the optimized transmission probabilities for the setting in Fig.~\ref{fig:opt_pi_eta.005} for the case of decoding without and with capture, respectively, for a high update generation rate with $\alpha \Ur = 2.5$. 
		The optimized transmission probabilities differ across the selected metrics, although average-AoI optimal and \gls{AVP} optimal probabilities are similar. 
		Without capture, the optimized probabilities indicate that a device should only transmit when it has enough energy, that is, when the single-user decoding error probability is low. Furthermore, the device should put aside some energy if its battery level is high, so that it can transmit again if a new update arrives shortly afterwards. 
		On the contrary, with capture, the devices should transmit with either high energy (using all $8$ energy units if they have a full battery) or moderate energy (i.e., using $3$ or $4$ energy units). The resulting variation in the energy of the packets facilitates 
		\gls{SIC}. 
		Note that the optimized transmission probabilities for the \gls{AoI} metrics are close to those of a strategy in which the devices either remain silent or transmit with all available energy, as considered in~\cite{Hoang2023Globecom}. Indeed, comparing Fig.~\ref{fig:opt_pi_eta.005} and~\cite[Fig.~5]{Hoang2023Globecom}, we see that the gains achievable by further tuning the packet energy level are marginal.
		
		\subsection{Impact of Battery Capacity} \label{sec:result_impact_E}
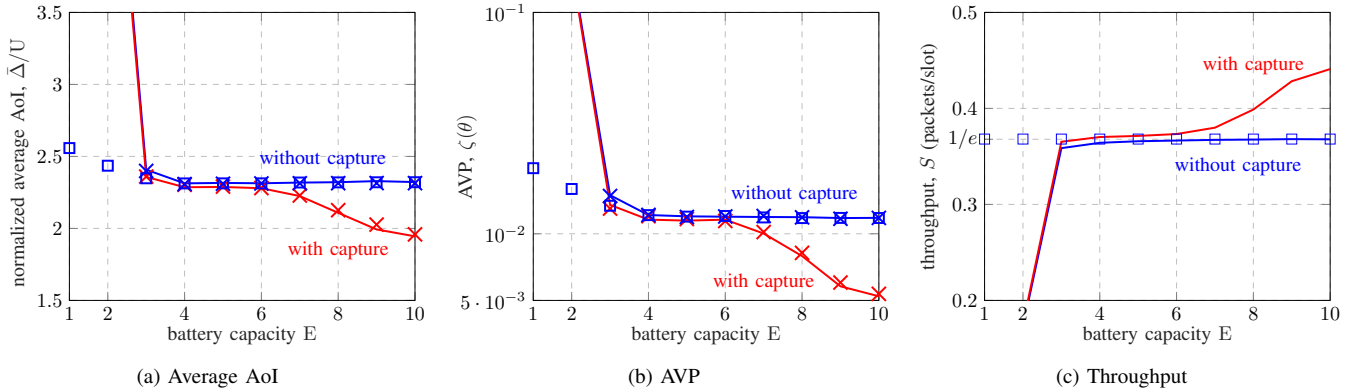
\begin{figure*}[t!]
			\centering
			\subcaptionbox{Average AoI \label{fig:avgAoI}
}{
	\begin{tikzpicture}[scale=.75]
		\begin{axis}[%
			width=2.4in,
			height=2in,	
			scale only axis,
			unbounded coords=jump,
			xmin=1,
			xmax=10,
			xtick= {1, 2, 4, 6,8,10},
			xlabel style={font=\color{white!15!black},yshift=4pt},
			xlabel={battery capacity $\Er$},
			ymin=1.5,
			ymax=3.5,
			yminorticks=true,
			ylabel style={font=\color{white!15!black},yshift=-3pt,xshift=-3pt},
			ylabel={normalized average AoI, $\bar{\Delta}/\Ur$},
			axis background/.style={fill=white},
			title style={font=\bfseries},
			xmajorgrids,
			ymajorgrids,
			yminorgrids,
			legend style={at={(.99,1)}, anchor=north east, legend cell align=left, align=left,draw=none, fill=white, fill opacity=.8,text opacity = 1}
			]

            \addplot [mark = x,color=red,only marks,mark size = 5,forget plot,line width = 1]
            table[row sep=crcr]{%
				3.0e+00 2.3598e+00 \\ 
4.0e+00 2.3028e+00 \\ 
5.0e+00 2.2905e+00 \\ 
6.0e+00 2.2803e+00 \\ 
7.0e+00 2.2285e+00 \\ 
8.0e+00 2.1266e+00 \\ 
9.0e+00 2.0265e+00 \\ 
1.0e+01 1.9586e+00 \\ 
			};

            \addplot [mark = x,color=blue,only marks,mark size = 5,forget plot,line width = 1]
			table[row sep=crcr]{%
				3.0e+00 2.3992e+00 \\ 
4.0e+00 2.3079e+00 \\ 
5.0e+00 2.3101e+00 \\ 
6.0e+00 2.3121e+00 \\ 
7.0e+00 2.3133e+00 \\ 
8.0e+00 2.3141e+00 \\ 
9.0e+00 2.3149e+00 \\ 
1.0e+01 2.3144e+00 \\ 
			};
   
            \addplot [mark = square,color=blue,only marks,mark size = 2.5,forget plot,line width = 1]
			table[row sep=crcr]{%
				1.000000000000000   2.557934174748731 \\
   2.000000000000000   2.435468492442428 \\
   3.000000000000000   2.348027223692171 \\
   4.000000000000000   2.312535364927302 \\
   5.000000000000000   2.312955464898252 \\
   6.000000000000000   2.313000070255622 \\
   7.000000000000000   2.318681810199244 \\
   8.000000000000000   2.319655029914951 \\
   9.000000000000000   2.320456415502996 \\
  10.000000000000000   2.320883007695036 \\
			};

            \addplot [line width = 1,color=blue,forget plot]
			table[row sep=crcr]{%
				2.0e+00 5.5148e+0 \\ 
				3.0e+00 2.4078e+0 \\ 
				4.0e+00 2.3130e+0 \\ 
				5.0e+00 2.3159e+0 \\ 
				6.0e+00 2.3135e+0 \\ 
				7.0e+00 2.3182e+0 \\ 
				8.0e+00 2.3210e+0 \\ 
				9.0e+00 2.3285e+0 \\ 
				1.0e+01 2.3218e+0 \\ 
			};
			
			\addplot [line width = 1,color=red,forget plot]
			table[row sep=crcr]{%
				2.0e+00 5.4287e+00 \\ 
				3.0e+00 2.3576e+00 \\ 
				4.0e+00 2.2861e+00 \\ 
				5.0e+00 2.2878e+00 \\ 
				6.0e+00 2.2791e+00 \\ 
				7.0e+00 2.2246e+00 \\ 
				8.0e+00 2.1084e+00 \\ 
				9.0e+00 1.9929e+00 \\ 
				1.0e+01 1.9449e+00 \\ 
			};

			\node[align = center,color=red] at (axis cs:8,1.85) () {with capture};
			\node[align=center,color=blue] at (axis cs:7.6,2.48) () {without capture};
			
		\end{axis}
\end{tikzpicture}}
\subcaptionbox{\gls{AVP} \label{fig:AVP}
}{
	\begin{tikzpicture}[scale=.75]
		\begin{axis}[%
			width=2.4in,
			height=2in,
			scale only axis,
			unbounded coords=jump,
			xmin=1,
			xmax=10,
			xtick= {1, 2, 4, 6,8,10},
		xlabel style={font=\color{white!15!black},yshift=4pt},
		xlabel={battery capacity $\Er$},
			ymode=log,
			ymin=5e-3,
			ymax=.1,
			ytick= {5e-3, 1e-2, 1e-1,1},
			yticklabels={$5\cdot10^{-3}$,$10^{-2}$,$10^{-1}$},
			yminorticks=true,
			ylabel style={font=\color{white!15!black},yshift=-15pt},
			ylabel={\gls{AVP}, $\zeta(\theta)$},
			axis background/.style={fill=white},
			title style={font=\bfseries},
			xmajorgrids,
			ymajorgrids,
			legend style={at={(.99,1)}, anchor=north east, legend cell align=left, align=left,  fill=white, fill opacity=.8,text opacity = 1,draw=none}
			]

            \addplot [mark = x,color=red,only marks,mark size = 5,forget plot,line width=1]
			table[row sep=crcr]{%
				3.0e+00 1.3000e-02 \\ 
4.0e+00 1.2010e-02 \\ 
5.0e+00 1.1663e-02 \\ 
6.0e+00 1.1422e-02 \\ 
7.0e+00 1.0182e-02 \\ 
8.0e+00 8.1992e-03 \\ 
9.0e+00 6.0223e-03 \\ 
1.0e+01 5.3538e-03 \\ 
			};

            \addplot [mark = x,color=blue,only marks,mark size = 5,forget plot,line width=1]
			table[row sep=crcr]{%
				3.0e+00 1.4818e-02 \\ 
4.0e+00 1.2171e-02 \\ 
5.0e+00 1.1912e-02 \\ 
6.0e+00 1.1971e-02 \\ 
7.0e+00 1.2037e-02 \\ 
8.0e+00 1.1864e-02 \\ 
9.0e+00 1.1691e-02 \\ 
1.0e+01 1.1796e-02 \\ 
			};
   
            \addplot [mark = square,color=blue,only marks,mark size = 2.5,forget plot,line width=1]
			table[row sep=crcr]{%
				1.000000000000000   0.019790368694667 \\
   2.000000000000000   0.015939978771699 \\
   3.000000000000000   0.013384232228090 \\
   4.000000000000000   0.012143191058474 \\
   5.000000000000000   0.011952755924718 \\
   6.000000000000000   0.012037598479408 \\
   7.000000000000000   0.011853751713126 \\
   8.000000000000000   0.011775715271627 \\
   9.000000000000000   0.011795250891224 \\
  10.000000000000000   0.011783068444850 \\
			};
   
			\addplot [line width = 1, color=blue,forget plot]
			table[row sep=crcr]{%
				2.0e+00 1.6285e-01 \\ 
				3.0e+00 1.4945e-02 \\ 
				4.0e+00 1.2153e-02 \\ 
				5.0e+00 1.1991e-02 \\ 
				6.0e+00 1.1945e-02 \\ 
				7.0e+00 1.1909e-02 \\ 
				8.0e+00 1.1885e-02 \\ 
				9.0e+00 1.1779e-02 \\ 
				1.0e+01 1.1800e-02 \\ 
			};
			
			\addplot [line width = 1, color=red,forget plot]
			table[row sep=crcr]{%
2.0e+00 1.5824e-01 \\ 
3.0e+00 1.3504e-02 \\ 
4.0e+00 1.1588e-02 \\ 
5.0e+00 1.1468e-02 \\ 
6.0e+00 1.1583e-02 \\ 
7.0e+00 1.0027e-02 \\ 
8.0e+00 7.9338e-03 \\ 
9.0e+00 5.7904e-03 \\ 
1.0e+01 5.2206e-03 \\ 
			};
			
			\node[align = center,color=red] at (axis cs:7,6e-3) () {with capture};
			\node[align=center,color=blue] at (axis cs:7.8,1.5e-2) () {without capture};
		\end{axis}
\end{tikzpicture}}
\subcaptionbox{Throughput \label{fig:throughput}
}{
	\begin{tikzpicture}[scale=.75]
		\begin{axis}[%
			width=2.4in,
			height=2in,
			scale only axis,
			unbounded coords=jump,
			xmin=1,
			xmax=10,
			xtick= {1, 2, 4, 6,8,10},
			xlabel style={font=\color{white!15!black},yshift=4pt},
			xlabel={battery capacity $\Er$},
			ymin = 0.2,
			ymax =.5,
			ytick = {0, 0.1, 0.2, 0.3, 0.367879441171442, 0.4,.5},
            yticklabels = {$0$, $0.1$, $0.2$, $0.3$, $1/e$, $0.4$,$0.5$},
			yminorticks=true,
			ylabel style={font=\color{white!15!black},yshift=-2pt},
			ylabel={throughput, $S$ (packets/slot)},
			axis background/.style={fill=white},
			title style={font=\bfseries},
			xmajorgrids,
			ymajorgrids,
			legend style={at={(.99,0.01)}, anchor=south east, legend cell align=left, align=left,draw=none, fill=white, fill opacity=.8,text opacity = 1}
			]

           \addplot [mark = square,color=blue,only marks,mark size = 2.5,forget plot]
        			table[row sep=crcr]{%
                    1.000000000000000   0.368063488259214 \\
                   2.000000000000000   0.368063488259315 \\
                   3.000000000000000   0.368063488259328 \\
                   4.000000000000000   0.368063488259339 \\
                   5.000000000000000   0.368063488259340 \\
                   6.000000000000000   0.368063488259337 \\
                   7.000000000000000   0.368063488259374 \\
                   8.000000000000000   0.368063488259388 \\
                   9.000000000000000   0.368063488259374 \\
                  10.000000000000000   0.368063488259385 \\
        			};
           
			\addplot [line width = 1, color=blue,forget plot]
			table[row sep=crcr]{%
				2.0e+00 1.7256e-01 \\ 
				3.0e+00 3.5862e-01 \\ 
				4.0e+00 3.6417e-01 \\ 
				5.0e+00 3.6581e-01 \\ 
				6.0e+00 3.6653e-01 \\ 
				7.0e+00 3.6723e-01 \\ 
				8.0e+00 3.6754e-01 \\ 
				9.0e+00 3.6800e-01 \\ 
				1.0e+01 3.6779e-01 \\ 
			};
			
			\addplot [line width = 1, color=red,forget plot]
			table[row sep=crcr]{%
				2.0e+00 1.7562e-01 \\ 
				3.0e+00 3.6527e-01 \\ 
				4.0e+00 3.7019e-01 \\ 
				5.0e+00 3.7135e-01 \\ 
				6.0e+00 3.7331e-01 \\ 
				7.0e+00 3.7994e-01 \\ 
				8.0e+00 3.9881e-01 \\ 
				9.0e+00 4.2832e-01 \\ 
				1.0e+01 4.4099e-01 \\ 
			};
			
			\node[align = center,color=red] at (axis cs:8,.445) () {with capture};
			\node[align=center,color=blue] at (axis cs:7.6,.34) () {without capture};
		\end{axis}
\end{tikzpicture}}
			\caption{The minimized average AoI, minimized \gls{AVP}, and maximized throughput vs. the battery capacity $\Er$. 
				Here, $\Ur = 1000$, $\alpha \Ur = 2$, $\EHrate = 0.005$, $\Nr = 100$, $\Rr = 0.8$, $\theta = 10^4$, and $\sigma^2 = -20$~dB. \revise{The solid lines and cross markers represent approximation and simulation results, respectively.} The square markers represent the performance assuming no noise, i.e., $\sigma^2 = 0$.}
			\label{fig:metrics_vs_E}
		\end{figure*}

		In Fig.~\ref{fig:metrics_vs_E}, we plot the average AoI, \gls{AVP}, and throughput for the optimal transmission strategy, as a function of the battery capacity~$\Er$. We consider a setting similar to Fig.~\ref{fig:opt_pi_eta.005}, except that we fix~$\alpha \Ur$ to $2$ and vary~$\Er$ from $2$ to $10$. We observe that, without capture, the AoI metrics and throughput do not improve when $\Er$ exceeds $4$. Indeed, when $\Er \ge 4$, the main cause of error is packet collision and not transmission errors due to noise. This can be seen by observing the performance achieved for $\sigma^2 = 0$ (i.e., for a collision channel), also depicted in Fig.~\ref{fig:metrics_vs_E}. Hence, further increasing $\Er$, which results in a larger value for $\bt$, is not helpful. 
		On the contrary, with capture, the AoI metrics and throughput keep on improving as the devices can store more energy. This is because transmitting with a higher energy $\bt$ facilitates 
		\gls{SIC}.
		
		\subsection{Impact of Energy Harvesting Rate} \label{sec:result_impact_eta}
\begin{figure*}[t!]
			\centering
			\input{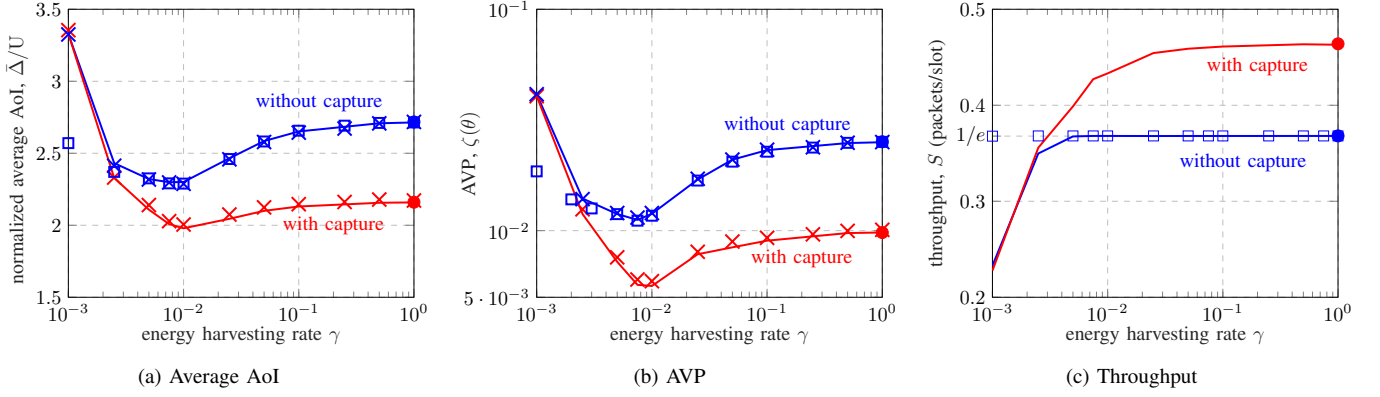}
			\caption{The minimized average AoI, minimized \gls{AVP}, and maximized throughput vs. the energy harvesting rate $\EHrate$.
				Here, $\Ur = 1000$, $\alpha \Ur = 2$, $\Er = 8$, $\Nr = 100$, $\Rr = 0.8$, $\theta = 10^4$, and $\sigma^2 = -20$~dB. The circle markers represent the performance assuming that the devices have always-full battery. \revise{The solid lines and cross markers represent approximation and simulation results, respectively.} The square markers represent the performance assuming no noise, i.e., $\sigma^2 = 0$.}
			\label{fig:metrics_vs_eta}
		\end{figure*}

		In Fig.~\ref{fig:metrics_vs_eta}, we plot the average AoI, \gls{AVP}, and throughput achieved with the optimal transmission strategy, as a function of the energy harvesting rate $\EHrate$. We consider a setting similar to Fig.~\ref{fig:opt_pi_eta.005}, except that we fix $\alpha \Ur$ to~$2$ and vary $\EHrate$ from $10^{-3}$ to~$1$. We also depict the performance achieved for $\sigma^2 = 0$, which shows that the impact of noise becomes apparent when $\EHrate$ is small. As $\EHrate \to 1$, the AoI metrics and throughput approach the performance for the case where the devices have always-full battery, which we analyzed in Section~\ref{sec:fullBat_slotedALOHA}. Remarkably, while the throughput increases as the energy harvesting rate grows, the average \gls{AoI} and \gls{AVP} are minimized at a value $\EHrate$ of around~$0.01$ and then increase with $\EHrate$. This is explained as follows. For a small $\EHrate$, the lack of energy forces the devices to transmit infrequent updates. 
		For a large $\EHrate$, the devices often have enough energy and transmit regardless of the obtainable age reduction, leading to many transmissions that cause collisions and, even if successful, result in a small \gls{AoI} reduction. 
		A moderate~$\EHrate$ 
		naturally sets appropriate thresholds such that updates with higher age reduction are transmitted with higher probability. 
		
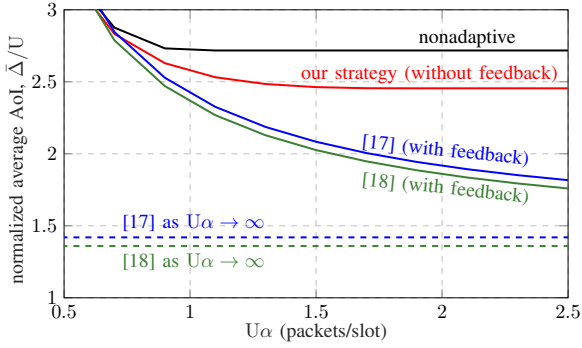
\begin{figure}[t!]
			\centering
			\begin{tikzpicture}[scale=.75]
		\begin{axis}[%
			width=3.5in,
			height=2in,	
			scale only axis,
			unbounded coords=jump,
			xmin=.5,
			xmax=2.5,
			xtick= {0, 0.5, 1, 1.5,2,2.5},
			xlabel style={font=\color{white!15!black},yshift=4pt},
			xlabel={$\Ur \alpha$ (packets/slot)},
			ymin=1,
			ymax=3,
			yminorticks=true,
			ylabel style={font=\color{white!15!black},yshift=-5pt,xshift=-3pt},
			ylabel={normalized average AoI, $\bar{\Delta}/\Ur$},
			axis background/.style={fill=white},
			title style={font=\bfseries},
			xmajorgrids,
			ymajorgrids,
			yminorgrids,
			legend style={at={(.99,1)}, anchor=north east, legend cell align=left, align=left,draw=none, fill=white, fill opacity=.8,text opacity = 1}
			]
   
			\addplot [line width = 1,color=black,forget plot]
			table[row sep=crcr]{%
				0.300000000000000   4.498881962906291 \\
   0.500000000000000   3.296705957500394 \\
   0.700000000000000   2.875980564693930 \\
   0.900000000000000   2.732039456051997 \\
   1.100000000000000   2.717422574230669 \\
   1.300000000000000   2.717422574262443 \\
   1.500000000000000   2.717422574332391 \\
   1.700000000000000   2.717422574227449 \\
   1.900000000000000   2.717422574229510 \\
   2.100000000000000   2.717422574226934 \\
   2.300000000000000   2.717422574240391 \\
   2.500000000000000   2.717422574529761 \\
			};

            \addplot [line width = 1,color=red,forget plot]
			table[row sep=crcr]{%
				0.300000000000000   4.496462500303108 \\
   0.500000000000000   3.282951021608533 \\
   0.700000000000000   2.832139709551722 \\
   0.900000000000000   2.629289202019389 \\
   1.100000000000000   2.531619803082731 \\
   1.300000000000000   2.484373727635146 \\
   1.500000000000000   2.462994567093848 \\
   1.700000000000000   2.455441528798734 \\
   1.900000000000000   2.454757115069282 \\
   2.100000000000000   2.454757115079026 \\
   2.300000000000000   2.454757115108808 \\
   2.500000000000000   2.454757115086378 \\
			};

            \addplot [line width = 1,color=blue,forget plot]
			table[row sep=crcr]{%
				0.300000000000000   4.750233333333334 \\
   0.500000000000000   3.416900000000000 \\
   0.700000000000000   2.845471428571429 \\
   0.900000000000000   2.528011111111111 \\
   1.100000000000000   2.325990909090909 \\
   1.300000000000000   2.186130769230769 \\
   1.500000000000000   2.083566666666667 \\
   1.700000000000000   2.005135294117647 \\
   1.900000000000000   1.943215789473684 \\
   2.100000000000000   1.893090476190476 \\
   2.300000000000000   1.851682608695652 \\
   2.500000000000000   1.816900000000000 \\
			};

            \addplot [line width = 1,color=OliveGreen,forget plot]
			table[row sep=crcr]{%
				0.300000000000000   4.692474247562856 \\
   0.500000000000000   3.359140914229522 \\
   0.700000000000000   2.787712342800951 \\
   0.900000000000000   2.470252025340633 \\
   1.100000000000000   2.268231823320431 \\
   1.300000000000000   2.128371683460291 \\
   1.500000000000000   2.025807580896189 \\
   1.700000000000000   1.947376208347170 \\
   1.900000000000000   1.885456703703207 \\
   2.100000000000000   1.835331390419999 \\
   2.300000000000000   1.793923522925175 \\
   2.500000000000000   1.759140914229523 \\
			};

            \addplot [line width = 1,color=OliveGreen,dashed,forget plot]
			table[row sep=crcr]{%
				0.3 1.359140914229523 \\   
                2.5 1.359140914229523 \\  
			};

            \addplot [line width = 1,color=blue,dashed,forget plot]
			table[row sep=crcr]{%
				0.3 1.4196 \\   
                2.5 1.4196 \\  
			};
			
			\node[align = center, color= black] at (axis cs:2.1,2.8) () {nonadaptive};
            \node[align = center, color = red] at (axis cs:1.95,2.55) () {our strategy (without feedback)};
            \node[align = center, color = blue,rotate = -6] at (axis cs:2,2.02) () {\cite{Yavascan2021} (with feedback)};
            \node[align = center, color = OliveGreen,rotate = -6] at (axis cs:2,1.76) () {\cite{Chen2022AoI} (with feedback)};
            \node[align = center, color = OliveGreen] at (axis cs:1,1.25) () {\cite{Chen2022AoI} as $\Ur\alpha \to\infty$};
            \node[align = center, color = blue] at (axis cs:1,1.52) () {\cite{Yavascan2021} as $\Ur\alpha \to\infty$};
		\end{axis}
        \end{tikzpicture}
			\caption{The normalized average \gls{AoI} $\bar{\Delta}/\Ur$ vs. average total number of new updates in a slot $\Ur\alpha$ for the case of always-full unit-sized battery. Here, $\Ur = 1000$, $\sigma^2 = 0$, and the decoding is without capture (collision channel).}
			\label{fig:adaptive_strategies}
		\end{figure}

		Note that the observed detrimental effect of a high $\EHrate$ is due to the considered slotted ALOHA protocol with fixed transmission probabilities across slots. As pointed out in     
		Section~\ref{sec:fullBat_thresholdALOHA}, if the devices have always-full battery, the same performance as energy-harvesting slotted ALOHA can be achieved for every $\EHrate$ provided that a multi-threshold strategy is used, where the thresholds are drawn independently from a ${\rm Geo}(\EHrate)$ distribution. This means that Fig.~\ref{fig:metrics_vs_eta} also depicts the performance of multi-threshold slotted ALOHA, as a function of the thresholds' parameter $\EHrate$. This figure highlights the need to adapt the transmission probability to the state of the devices in each slot. In~\cite{Yavascan2021, Chen2022AoI} the benefits of this adaptation are demonstrated for the case in which feedback from the receiver is available. Fig.~\ref{fig:metrics_vs_eta} demonstrates that this adaptation is beneficial also when there is no feedback. To further highlight this benefit, in Fig~\ref{fig:adaptive_strategies}, we consider a collision channel and plot the normalized average \gls{AoI} $\bar{\Delta}/\Ur$ achieved with the nonadaptive scheme (Theorem~\ref{th:approx_AoI_fullBat}), our multi-threshold slotted ALOHA scheme with optimized $\EHrate$, and the adaptive strategies with feedback proposed in~\cite{Yavascan2021,Chen2022AoI}. When $\Ur$ is large, the strategies in \cite{Yavascan2021} and \cite{Chen2022AoI} achieve a $\bar{\Delta}/\Ur$ of $1.4196 + 1/(\Ur \alpha)$ and $e/2 + 1/(\Ur \alpha)$, respectively (see \cite[Sec.~IV]{Yavascan2021}). While our adaptive strategy has a clear advantage over the nonadaptive one, the question of optimally designing an adaptive strategy without feedback 
		remains open.
		
		\section{Conclusions} \label{sec:conclusions}
		We studied the impact of energy harvesting on information freshness in slotted ALOHA networks. Leaning on a Markovian analysis, we provided an exact  analysis of the average AoI, as well as 
		easy-to-compute and accurate approximations for both the average AoI and the \gls{AVP}. 
		We showed that transmitting a new update whenever possible is beneficial only for low update generation rates, while waiting for sufficient energy before transmitting is preferable for high update generation rates. Significant gains with respect to these two baselines can be achieved with an optimized strategy. The AVP-minimizing strategy performs well also in terms of
		the average AoI and vice versa. However, the throughput-maximizing strategy entails a notable loss in terms of the AoI metrics when the update generation rate is high. Decoding with capture significantly outperforms decoding without capture. 
		Our results also highlight the benefit of adapting the transmission probability with respect to both the battery level and the time elapsed since the last transmission of a device. 
		
	
	\appendix

	\revise{
		\subsection{On the System Model}
		We provide below some remarks about our system model and discuss possible extensions.
		
		\subsubsection{Energy Harvesting Model}
		The time-uncorrelated Bernoulli model in the paper captures the randomness of energy harvesting while allowing for tractable analysis. In practice, the energy harvesting processes can be time-correlated, 
		and models for such case can be found in~\cite[Sec.~II-B]{Ku16_energyHarvesting}. For example, one can model the energy-harvesting process as a two-state Markov chain consisting of a ``good'' state with a high energy-harvesting rate and a ``bad'' state with a low energy-harvesting rate~\cite{Michelusi13}. These states represent periods with abundant and scarce ambient energy, respectively. In such a scenario, the devices can adapt the transmission probabilities to their energy harvesting state. To do so, one would need to extend the Markov chain $G\of{s}$ in Section~\ref{sec:Markov_G} to capture the energy harvesting state of the devices and follow the same method in this paper to analyze the \gls{AoI}.
		
		\subsubsection{Homogeneous vs. Heterogeneous Settings} 
		For simplicity, we assume homogeneous devices with the same battery capacity $\Er$, update generation rate $\alpha$, energy harvesting rate $\EHrate$, and transmission probabilities $\{\pi_{b,\bt}\}$. Therefore, the devices are stochastically equivalent, allowing us to focus on a generic device. The analysis can be extended to heterogeneous devices belonging to multiple classes with different parameters and requirements, as considered in~\cite{Hoang2021AoI}. 
		In this case, the AoI refresh status $X\of{s}$, battery level $B\of{s}$, and battery profile $\Lm\of{s}$ become class-dependent. The Markov chain $G\of{s}$ should be extended accordingly. The exact and approximate AoI analyses follow the same mathematical machinery in Sections~\ref{sec:AoI} and~\ref{sec:AoI_approx}, respectively. The accuracy of the approximation proposed in Section~\ref{sec:AoI_approx} may deteriorate with increased heterogeneity.
		
		\subsubsection{Single vs. Multiple Transmission Attempts} 
		We assume a single transmission attempt per update. Our analysis can be easily extended to the case where the devices retransmit the latest update whenever they do not have a new one, as considered in~\cite{Munari2022_retransmission}. 
		However, our experiments show no apparent \gls{AoI} reduction by retransmissions. Therefore, we focus only on the case without retransmissions in this paper.
		
		\subsubsection{Battery Level Discretization} 
		We allow a device with battery level $b$ to select transmit power only from the finite set $[0:b]$. This discretization enables finite-state Markovian analysis and gives the devices a simple rule to adjust their transmit power. Note that, in low-complexity IoT devices, three power levels are typically considered (zero power, low power, and high power).
		
		\subsubsection{The Physical-Layer Channel Model}
		An extension to more complicated channel models than the \gls{AWGN} channel considered in the paper is straightforward, provided that one can compute the successful-decoding probability $\omega_{\bt,\Lm}$. This entails replacing the normal approximation~\eqref{eq:rate_FBL} with other appropriate formulas. For example, normal approximations of the achievable rate for the Rayleigh block fading channel can be found in~\cite{Lancho20SISOblockfading,Qi20MIMOblockfading}. Note that with fading, interference cancellation requires accurate channel estimation. Therefore, one may need to account for interference cancellation errors and residual interference.
	}
	
	\subsection{Proof of Theorem~\ref{th:AoI_YZ}} \label{proof:AoI_YZ}
	Without loss of generality, we start tracking the process (i.e., we set $t = 0$) right after the first \gls{AoI} refresh, which we index as the $0$th refresh. Let $t_i$ be the time instant of the $i$th \gls{AoI} refresh and $s_i$ be the slot corresponding to $t_i$. Let also $y_i = t_i - t_{i-1} = s_i-s_{i-1}$ be the duration of the $i$th inter-refresh period. We establish~\eqref{eq:PMF_peak_AoI} simply by noting that the peak \gls{AoI} is given by $\widetilde\Delta(i) = y_i + 1$. To compute the  \gls{PMF} of the discretized \gls{AoI}, we proceed as follows:
	\begin{align}
		\!\!\P{\widehat \Delta(s) \!=\! \delta}  
		&= \lim_{n \to\infty} \frac{1}{n} \sum_{s = 1}^n \ind{\widehat{\Delta}(s) = \delta} \\
		&= \lim_{m \to\infty} \frac{1}{\sum_{i=1}^m y_i} \sum_{i=1}^m\sum_{s = s_{i-1} + 1}^{s_i} \!\!\ind{\widehat{\Delta}(s) \!=\! \delta} \label{eq:tmp1390}\\
		&= \lim_{m \to\infty} \frac{1}{\sum_{i=1}^m y_i} \sum_{i=1}^m \ind{y_i+1 \ge \delta} \label{eq:tmp1392} \\
		&= \lim_{m \to \infty} \frac{1}{\frac{1}{m}\sum_{i=1}^m y_i}  \frac{|\{i \!\in\! [m] \colon y_i  \ge \delta \!-\! 1\}|}{m}\!  \\
		&= \frac{1}{\E{Y}} \P{Y \ge \delta-1}. \label{eq:tmp1393}
	\end{align}
	Here,~\eqref{eq:tmp1392} holds since within the $i$th inter-refresh period, $\widehat{\Delta}(s)$ increases linearly from $2$ to $y_i+1$, and thus there exists one slot where the AoI value is $\delta$ if and only if $y_i + 1 \ge \delta$; \eqref{eq:tmp1393} holds because $\frac{1}{m}\sum_{i=1}^m y_i \to \E{Y}$ and $\frac{|\{i \in [m] \colon y_i  \ge \delta-1\}|}{m} \to \P{Y\ge \delta-1}$ as $m \to \infty$. 
	
	The \gls{AoI} metric $F(\Delta) =  \lim\limits_{\bar{t} \to \infty} \frac{1}{\bar{t}}\int_{0}^{\bar t} f(\Delta(t))  {\rm d} t$ is derived as
	\begin{align}
		\!\!F(\Delta) 
		&= \lim_{m \to \infty} \frac{1}{\sum_{i=1}^m y_i} \sum_{i=1}^m \int_{t_{i-1}}^{t_i} f(\Delta(t)) {\rm d} t \\
		&= \lim_{m \to \infty} \frac{1}{\sum_{i=1}^m y_i} \sum_{i=1}^m \int_{1}^{y_i+1} f(t) {\rm d} t  \label{eq:tmp1612}\\
		&= \lim_{m \to \infty} \!\frac{1}{\frac{1}{m}\sum_{i=1}^m y_i} \sum_{y=0}^\infty \frac{|\{i \!\in\! [m]\colon y_i \!=\! y\}|}{m} \!\int_{1}^{y+1} \!\!\!f(t) {\rm d} t \notag  \\
		&= \frac{1}{\E{Y}} \E{\int_{1}^{Y+1} f(t) {\rm d} t}. \label{eq:tmp1403} 
	\end{align}
	Here,~\eqref{eq:tmp1612} follows by noting that within the $i$th inter-refresh period, $\Delta(t) = t - t_{i-1} + 1$ and by applying a change of variable; \eqref{eq:tmp1403} holds because $\frac{1}{m}\sum_{i=1}^m y_i \to \E{Y}$ and $\frac{|\{i \in [m]\colon y_i = y\}|}{m} \to \P{Y = y}$ as $m\to \infty$.
	For the average \gls{AoI}, $f(t) = t$, and we have that  $\int_1^{Y+1} f(t) {\rm d} t = Y + \frac{Y^2}{2}$. Substituting this into~\eqref{eq:tmp1403}, we obtain~\eqref{eq:avgAoI_YZ}. 
%
For the \gls{AVP}, $f(t) = \ind{t>\theta}$, and we have that $\int_1^{Y+1} f(t) {\rm d}t = (Y-\theta + 1)^+$. Indeed, within an inter-refresh period of duration $Y$, the AoI exceeds $\theta$ in the last $(Y-\theta+1)^+$ slots (see Fig.~\ref{fig:AoI_process}). Substituting this into~\eqref{eq:tmp1403}, we obtain
\begin{align}
	\!\!\!\zeta(\theta) 
	&= \frac{1}{\E{Y}} \sum_{y=0}^\infty \P{Y=y}(y - \theta + 1)^+ \label{eq:tmp649}\\
	&= \frac{1}{\E{Y}} \bigg(\sum_{y=\theta}^\infty y \P{Y=y} - (\theta\!-\!1) \sum_{y = \theta}^\infty \P{Y=y} \bigg) \\
	&= 1 - \frac{1}{\E{Y}} \bigg( \sum_{y=1}^{\theta-1} y \P{Y=y} - (\theta\!-\!1) \P{Y\ge \theta} \bigg). \label{eq:tmp392} 
\end{align}

\subsection{\revise{Terminating Markov Chain and}
	Discrete Phase-Type Distribution} \label{sec:phase_type}
In a Markov chain, an absorbing state is a state that, once entered, cannot be left. 
A Markov chain is called a {\em terminating Markov chain} if there is one absorbing state and it is possible to go from any non-absorbing state (called transient state) to the absorbing state in a finite number of steps~\cite[Chap.~III]{Kemeny1976_Markov}. 
By reordering the states, the transition probability matrix of a terminating Markov chain with $m$ transient states can be expressed as
$\Ps =\bigg[\begin{matrix}
	\Ts & \av \\ \mathbf{0}_{m}^\T & 1 
\end{matrix}\bigg]$,
where $\Ts \in [0,1]^{m \times m}$ contains the probabilities of transitions between transient states and $\av = (\Is_m - \Ts) \mathbf{1}_m$ contains the probabilities of transitions from the transient states to the absorbing states. 
The matrix $\Is_m - \Ts$ is invertible for every terminating Markov chain~\cite[Th.~3.2.1]{Kemeny1976_Markov}. 
%
The distribution of the absorption time, i.e., the number of steps until absorption, of a terminating Markov chain is called the {\it discrete phase-type distribution} and described next. 
\begin{lemma}[{Discrete phase-type distribution}] \label{lem:phase_type}
	Consider a terminating Markov chain with transition probability matrix $\Ps = \bigg[\begin{matrix}
		\Ts & \av \\ \mathbf{0}_m^\T & 1 
	\end{matrix}\bigg]$. 
	Let $Y$ be the absorption time when starting from the transient state $i\in [m]$ \gls{wp} $\tau_i$, i.e., the initial probability vector of the transient states is $\tauv = (\tau_1,  \tau_2, \dots, \tau_m)$. The \gls{PMF} and \gls{CCDF} of $Y$ are given by $\P{Y = y} = \tauv^\T \Ts^{y-1} \av$ and $\P{Y > y} = \tauv^\T \Ts^y \mathbf{1}_m$  for $y = 1, 2, \dots$. 
The first- and second-order moments of $Y$ are given by $\E{Y} = \tauv^\T (\Is_m - \Ts)^{-1}  \mathbf{1}_m$ and $\E{Y^2} = 2\tauv^\T (\Is_m - \Ts)^{-2}  \mathbf{1}_m - \E{Y}$. 
\end{lemma}
\begin{proof}
The \gls{PMF} of $Y$ is given in~\cite[Sec.~2.2]{Neuts1994}. 
The \gls{CCDF} of $Y$ can be obtained from its \gls{PMF} after some simple manipulations. 
As shown in~\cite[Sec.~2.2]{Neuts1994}, the factorial moments of $Y$ are $\E{\prod_{i=0}^{y-1}(Y-i)} = y! \tauv^\T (\Is_m \!-\! \Ts)^{-y} \Ts^{y-1} \mathbf{1}_m$, $y = 1,2,\dots$ By applying this result with $y=1$ and $y = 2$, we  obtain the first- and second-order moments of $Y$. 
\end{proof}

\revise{The discrete phase-type distribution has been used to model inter-arrival and service times in recent \gls{AoI} analyses, e.g.,~\cite{Akar20,Dogan21}.}

\section*{Acknowledgement}

K.-H. Ngo would like to thank Nikolaos Pappas for fruitful discussions.

\bibliographystyle{IEEEtran}
\bibliography{IEEEabrv,./biblio}

\begin{IEEEbiography}[{\includegraphics[width=1in,height=1.25in,clip,keepaspectratio]{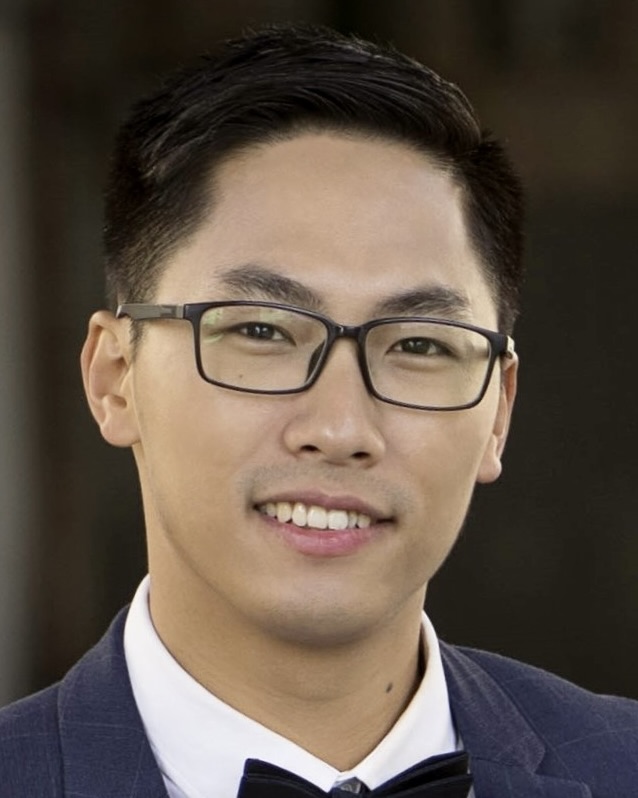}}]{Khac-Hoang Ngo}
(Member, IEEE) received the B.E. degree (Hons.) in electronics and telecommunications from University of Engineering and Technology, Vietnam National University, Hanoi, Vietnam, in 2014; and the M.Sc. degree (Hons.) and Ph.D. degree in wireless communications from CentraleSupélec, Paris-Saclay University, France, in 2016 and 2020, respectively. His Ph.D. thesis was also realized at Paris Research Center, Huawei Technologies France. Since September 2024, he has been an assistant professor with Linköping University, Sweden. From 2020 to 2024, he was a postdoctoral researcher with Chalmers University of Technology, Sweden. 
His research interests include wireless communications and information theory, with an emphasis on massive random access, privacy and security of machine learning, information freshness, MIMO, and noncoherent communications. He received the ``Signal, Image \& Vision Ph.D. Thesis Prize'' by Club EEA, GRETSI and GdR-ISIS, France in 2021.
\end{IEEEbiography}

\begin{IEEEbiography}[{\includegraphics[width=1in,height=1.25in,clip,keepaspectratio]{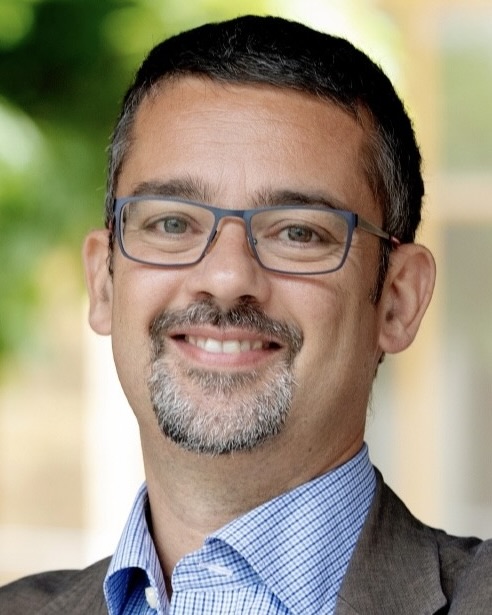}}]
{Giuseppe Durisi} (S'02-M'06-SM'12)
received the Laurea (summa cum laude) and Ph.D. degrees from the Politecnico di Torino, Italy,
in 2001 and 2006, respectively.  From 2002 to 2006, he was with the Istituto Superiore Mario
Boella, Turin, Italy. From 2006 to 2010, he was a Post-Doctoral Researcher at ETH Zurich,
Zürich, Switzerland. In 2010, he joined the Chalmers University of Technology, Gothenburg,
Sweden, where he is currently a Professor with the Communication Systems Group. His research
interests are in the areas of communication and information theory and machine learning. He is
the recipient of the 2013 IEEE ComSoc Best Young Researcher Award for the Europe, Middle East,
and Africa region, and is coauthor of a paper that won the Student Paper Award at the 2012
International Symposium on Information Theory, and of a paper that won the 2013 IEEE Sweden
VT-COM-IT Joint Chapter Best Student Conference Paper Award.  From 2015 to 2021, he served as
associate editor for the  IEEE Transactions on Communications.  Since 2024, he has been an
associate editor for the IEEE Transactions on Information Theory.
\end{IEEEbiography}

\begin{IEEEbiography}[{\includegraphics[width=1in,height=1.25in,clip,keepaspectratio]{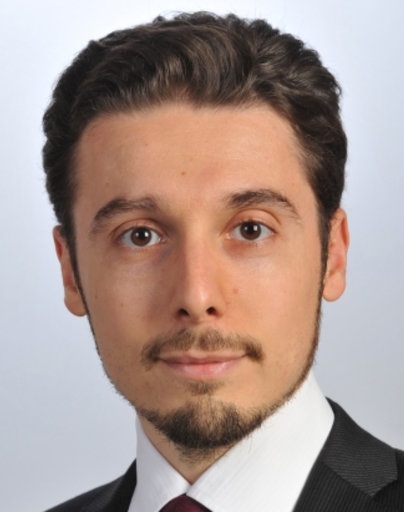}}]{Andrea Munari} (Senior Member, IEEE) received the M.Sc. and Ph.D. degrees in telecommunications engineering from the University of Padova, Padua, Italy, in 2006 and 2010, respectively. From 2007 to 2010, he was with IBM Research in Zurich, Switzerland. In 2011, he joined the Corporation Research and Development Division, Qualcomm Inc., San Diego, CA, US. He is currently with the Institute of Communications and Navigation, German Aerospace Center (DLR), Wessling, Germany. From 2014 to 2018, he held a Senior Researcher and a Lecturer position with the Institute of Networked Systems, RWTH Aachen University, Aachen, Germany. His main research interests are in the area of wireless communications, with special focus on massive IoT access and satellite communications. Dr. Munari received the 2018 ACM MobiCom Workshop on Millimeter Wave Networks and Sensing Systems Best Paper Award, and the IEEE Globecom 2020 Communications Theory Symposium Best Paper Award. He serves as Associate Editor of IEEE Communications Letters.
\end{IEEEbiography}

\begin{IEEEbiography}[{\includegraphics[width=1in,height=1.25in,clip,keepaspectratio]{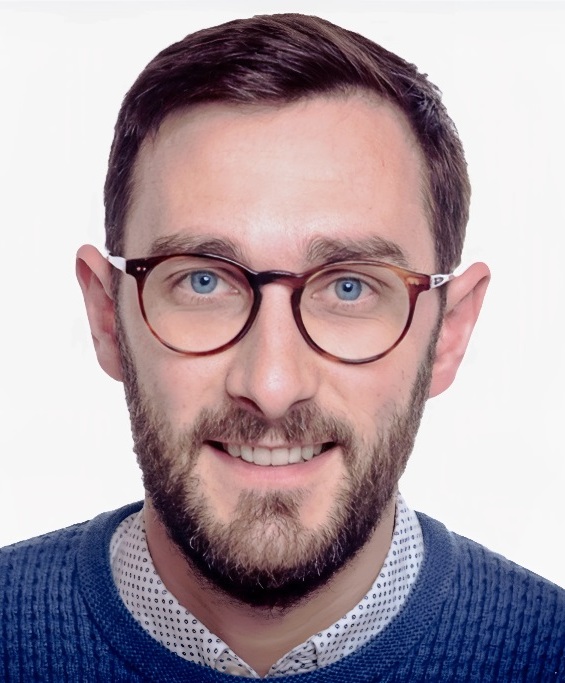}}]{Francisco Lázaro} (Senior Member, IEEE)
received his degree in telecommunications engineering from the University of Zaragoza, Spain, and his Ph.D. in electrical engineering from the Hamburg University of Technology, Germany, in 2006 and 2017, respectively. Since 2008, he has been part of the research staff at the Institute of Communications and Navigation of the German Aerospace Center (DLR), where he currently leads the Advanced Information Processing group. Since 2020, he has been a lecturer for satellite communications at the Karlsruhe Institute of Technology (KIT). His main research interests include multi-user communications and error-correcting codes.
\end{IEEEbiography}

\begin{IEEEbiography}[{\includegraphics[width=1in,height=1.25in,clip,keepaspectratio]{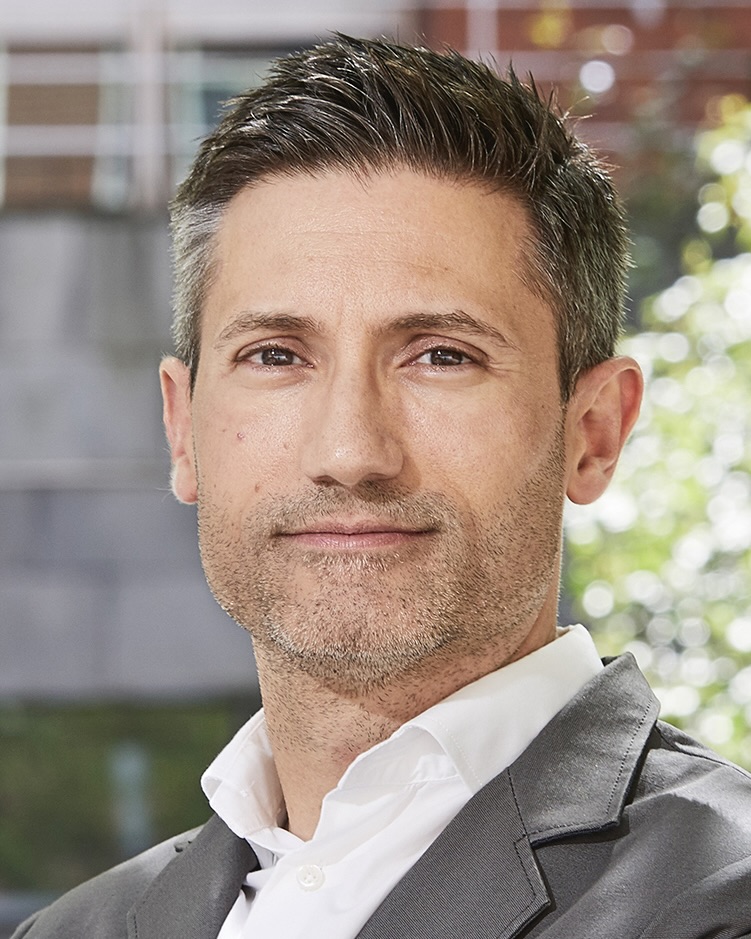}}]{Alexandre Graell i Amat}
(Senior Member, IEEE) received the M.Sc. and Ph.D. degrees in electrical engineering from the Politecnico di Torino, Turin,
Italy, in 2000 and 2004, respectively, and the M.Sc. degree in telecommunications engineering from the Universitat Politècnica de Catalunya, Barcelona, Catalonia, Spain, in 2001. From 2001 to 2002, he was a Visiting Scholar with the University of California at San Diego, La Jolla, CA,
USA. From 2002 to 2003, he held a visiting appointment at Universitat Pompeu Fabra, Barcelona, and the Telecommunications Technological Center of Catalonia, Barcelona. From 2001 to 2004, he held a part-time appointment
at STMicroelectronics Data Storage Division, Milan, Italy, as a Consultant on coding for magnetic recording channels. From 2004 to 2005, he was a Visiting Professor with Universitat Pompeu Fabra. From 2006 to 2010, he was with the Department of Electronics, IMT Atlantique (formerly ENST Bretagne), Brest, France. From 2019 to 2023, he was also been an Adjunct Research Scientist with Simula UiB, Bergen, Norway. He is currently a Professor with the Department of Electrical Engineering, Chalmers University of Technology,     Gothenburg, Sweden. His research interests are in the field of coding theory and AI security. He received the Marie Skłodowska-Curie Fellowship from the European Commission and the Juan de la Cierva Fellowship from the Spanish Ministry of Education and Science. He received the IEEE Communications Society 2010 Europe, Middle East, and Africa Region Outstanding Young Researcher Award. He was the General Co-Chair of the 7th International Symposium on Turbo Codes and Iterative Information Processing, Sweden, in 2012, and the TPC Co-Chair of the 11th International Symposium on Topics in Coding, Canada, in 2021. He was an Associate Editor of the IEEE COMMUNICATIONS LETTERS from 2011 to 2013. He was an Associate Editor and the Editor-at-Large of the IEEE TRANSACTIONS     ON COMMUNICATIONS from 2011 to 2016 and 2017 to 2020, respectively. He was an Area Editor of the IEEE TRANSACTIONS ON  COMMUNICATIONS from 2020 to 2023.
\end{IEEEbiography}

\end{document}